\documentclass[submission, Phys]{SciPost}

\pdfoutput=1
\usepackage{comment}
\usepackage{enumerate}
\usepackage{amssymb}
\usepackage{amsmath}
\usepackage{braket}
\usepackage{graphicx}
\usepackage{dsfont}

\newcommand{\be}{\begin{equation}}
\newcommand{\ee}{\end{equation}}
\newcommand{\ba}{\begin{aligned}}
\newcommand{\ea}{\end{aligned}}
\newcommand{\bw}{\begin{widetext}}
\newcommand{\ew}{\end{widetext}}

\renewcommand{\vec}[1]{\boldsymbol{#1}}
\newcommand{\bea}{\begin{eqnarray}}
\newcommand{\eea}{\end{eqnarray}}

\begin{document}

\begin{center}{\Large \textbf{
Logarithmic negativity in out-of-equilibrium open free-fermion chains: An exactly solvable case 
}}\end{center}

\begin{center}
Vincenzo Alba\textsuperscript{1*}
\end{center}

\begin{center}
{\bf 1}	
Dipartimento di Fisica dell' Universit\`a di Pisa and INFN, Sezione di Pisa, I-56127 Pisa, Italy\\
* vincenzo.alba@unipi.it
\end{center}

\begin{center}
Federico Carollo\textsuperscript{2}
\end{center}
\begin{center}
{\bf 2}
Institut f\"ur Theoretische Physik, Universit\"at T\"ubingen,
	Auf der Morgenstelle 14, 72076 T\"ubingen, Germany
\end{center}

\begin{center}
\today
\end{center}


\section*{Abstract}
{\bf
We derive the quasiparticle picture for the fermionic 
logarithmic negativity in a tight-binding chain subject to gain and loss dissipation. 
We focus on the dynamics after the quantum quench from 
the fermionic N\'eel state. 
We consider the negativity between both adjacent and disjoint intervals embedded in an 
infinite chain. Our result holds in the standard hydrodynamic limit of large subsystems  
and long times, with their ratio fixed. Additionally, 
we consider the weakly-dissipative limit, in which the dissipation rates are 
inversely proportional to the size of the intervals. 
We show that the negativity is proportional to the number of entangled pairs 
of quasiparticles that are shared between the two intervals, as is the case for the mutual 
information. Crucially,  in contrast with the unitary case, the negativity content of quasiparticles 
is not given by the R\'enyi entropy with R\'enyi index $1/2$, and it is in general not easily 
related to thermodynamic quantities. 
}

\section{Introduction}
\label{sec:intro}

Distinguishing genuine quantum correlations from statistical ones in quantum many-body systems 
is a daunting task. While for bipartite quantum systems in a pure state several \emph{computable} quantum 
information motivated measures can be used to identify entanglement~\cite{amico2008entanglement,eisert2010colloquium,calabrese2009entanglement,laflorencie2016quantum},  this is more challenging for mixed-state systems.  
Open quantum systems undergoing dissipative Lindblad dynamics~\cite{petruccione2002the,rossini2021coherent} 
represent an important example of systems featuring mixed states. Recently, it has been shown that for one-dimensional 
free-fermion and free-boson systems it is possible to describe the dynamics of 
information-related quantities, such as von Neumann and R\'enyi entropies, as well as the mutual 
information, in the presence of generic 
 quadratic dissipation~\cite{alba2021spreading,carollo2022dissipative,alba2022hydrodynamics}. 
This generalizes the well-known quasiparticle picture for 
entanglement spreading after quantum quenches in integrable 
systems~\cite{calabrese2005evolution,fagotti2008evolution,alba2017entanglement}. 
%
%
\begin{figure}[t]
\centering
\includegraphics[width=0.6\textwidth]{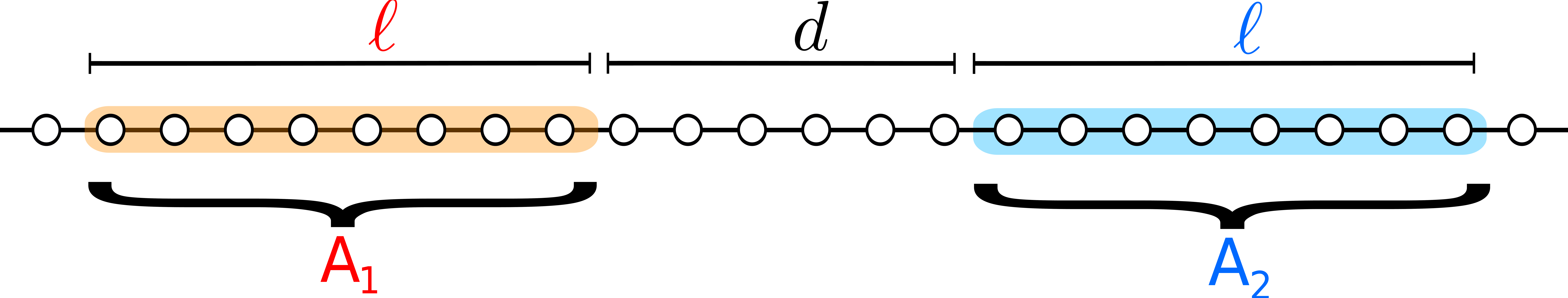}
\caption{ Logarithmic negativity $\cal{E}$ between two 
 intervals $A_1$ and $A_2$ of equal length $\ell$ at distance 
 $d$ embedded in an infinite chain. 
}
\label{fig0:bip}
\end{figure}
%
Although the R\'enyi entropies and the mutual information are not proper measures 
of entanglement for mixed states~\cite{plenio2007an}, it has been shown in Ref.~\cite{carollo2022dissipative} that 
even for Lindblad dynamics the mutual information is sensitive to the presence of 
correlated pairs of quasiparticles. 
This is similar to closed quantum systems, although the dissipation dramatically affects the correlation 
content of the quasiparticles. 

To make a step forward towards understanding entanglement dynamics in open quantum systems  
here we focus on the logarithmic negativity, which is a proper entanglement measure 
also for mixed states~\cite{lee2000partial,vidal2002computable,eisert2006entanglement,plenio2005logarithmic,calabrese2012entanglement}. 
The computation of the logarithmic negativity is in general a challeging task. 
It can be computed effectively from the two-point correlation function 
only for free-boson systems~\cite{audenaert2002entanglement}. For 
free fermionic ones it requires knowledge of the spectrum of the 
so-called partial transpose, which is not a Gaussian 
fermionic operator~\cite{eisler2015on}.  This means that the computational cost to 
extract the logarithmic negativity from a fermionic many-body wavefunction for a system with $L$ sites 
grows exponentially with $L$. Very recently, an alternative definition of 
negativity (that was dubbed \emph{fermionic} negativity) has been 
proposed~\cite{shapourian2019entanglement} for free fermions. This fermionic negativity can be computed 
from the two-point fermionic correlation functions.  The computational cost to determine the 
fermionic logarithmic negativity grows only polynomially with $L$.  
For this reason here we restrict ourselves to the fermionic logarithmic negativity. 
For generic  interacting systems, both 
the standard and the fermionic logarithmic negativity 
can be computed with Matrix Product States (MPS) methods (see Ref.~\cite{ruggiero2016entanglement} for 
the standard negativity), at least at 
equilibrium. Importantly, both the standard and the fermionic negativity are 
proper entanglement measures for mixed states (see Ref.~\cite{shapourian2019entanglement} for 
a careful comparison between them).

The negativity is attracting increasing attention as a tool to characterize 
universal aspects of equilibrium and out-of-equilibrium quantum many-body systems 
(see section~\ref{sec:negativity}). 

Interestingly, it has been shown 
in Ref.~\cite{alba2019quantum} that after quantum quenches in 
one-dimensional closed quantum integrable systems, both the standard negativity and the 
fermionic one become equal to half  the R\'enyi mutual information with R\'enyi index $1/2$. 
This has been verified for both free-fermion and free-boson models. For generic interacting systems 
it is quite challenging to build a quasiparticle picture to describe the full-time 
dynamics of R\'enyi entropies, although 
their value in the steady state can be determined~\cite{alba2017quench,alba2017renyi,mestyan2018renyi,alba2019towards}. 
Moreover, recent exact results for quenches in the so-called rule 
$54$ chain~\cite{klobas2021entanglement}, which is a ``minimal model'' 
for  interacting integrable systems, suggest that R\'enyi entropies violate 
the quasiparticle picture paradigm. These results motivated a 
conjecture for the growth with time of the R\'enyi entropies in 
generic interacting integrable systems~\cite{bertini2022growth}. 
However, Ref.~\cite{bertini2022entanglement} showed that in the early-time regime and for 
contiguous subsystems the relation between Renyi-$1/2$ mutual 
information and logarithmic negativity put forward in Ref.~\cite{alba2019quantum} 
still holds for  any local 
quantum circuit, and therefore also for interacting integrable 
systems. Similar results were obtained in CFTs~\cite{kudler2021the}.

In the context of open quantum systems subject to a Lindblad dynamics the logarithmic negativity has not been 
explored much. Some numerical results were presented in Ref.~\cite{alba2021spreading}, suggesting that 
in the presence of dissipation the  negativity is not half of the R\'enyi mutual information 
with R\'enyi index $1/2$, in contrast with closed systems~\cite{alba2019quantum}. 
Here we derive the quasiparticle picture for the  fermionic logarithmic negativity after the 
quench from the fermionic N\'eel state in a tight-binding chain with 
homogeneous gain and loss of fermions. We consider the geometry sketched in Fig.~\ref{fig0:bip}, 
focusing on the entanglement between two intervals  $A_1,A_2$ of length $\ell$ and placed at a distance $d$. The intervals  
are embedded in an infinite chain. 
Our results hold in the standard hydrodynamic limit of 
long times, large subsystem size, and large distances, i.e., $\ell,t,d\to\infty$, 
with the ratios $t/\ell$ and $d/\ell$ fixed and arbitrary. The logarithmic negativity decays 
expontially in time with a rate depending on the gain 
and loss rates $\gamma^\pm$. Since we consider times of order $\ell$, in order to observe a 
non-trivial time-evolution of the logarithmic negativity (and not an instantaneous convergence 
to its stationary value) we work in the weakly-dissipative 
hydrodynamic limit, obtained by taking vanishing $\gamma^\pm\to0$, with  
fixed $\gamma^\pm\ell$.  Our results show that the dynamics of the logarithmic negativity 
can be described within the framework of the quasiparticle picture. 
Specifically, we show that the logarithmic negativity, and hence the 
entanglement, is proportional to the number of 
entangled pairs of quasiparticles that are shared between the two intervals. Indeed, 
the structure of our formula for the logarithmic 
negativity is the same as that of the mutual information~\cite{carollo2022dissipative}. 
The contribution of the entangled quasiparticles 
to the negativity is time-dependent and it vanishes at long times, 
again, similar to the mutual information~\cite{carollo2022dissipative}. 
However, in contrast with the unitary case~\cite{alba2019quantum},  
this negativity content  does not coincide with 
the R\'enyi entropy with R\'enyi index $1/2$, and it is not, in general, straightforwardly related to known thermodynamic 
quantities. 

The manuscript is organized as follows. In section~\ref{sec:model} we introduce the tight-binding 
chain and discuss the treatment of the quench from the fermionic N\'eel state. In section~\ref{sec:lin} we 
review the Lindblad framework for gain and loss dissipation. 
In section~\ref{sec:quasi-diss} we discuss the quasiparticle picture for free systems with 
quadratic dissipation. In section~\ref{sec:negativity} we introduce the fermionic 
logarithmic negativity. As a 
warm-up, we present in section~\ref{sec:warm-up} an \emph{ab initio} derivation of the quasiparticle 
picture for the R\'enyi entropies~\cite{alba2021spreading}. This was obtained already 
in Ref.~\cite{alba2021spreading} using the results of Ref.~\cite{calabrese2012quantumquench},
although the alternative derivation that we present here is new and self-contained. In section~\ref{sec:neg-calculation} we 
derive the hydrodynamic picture for the logarithmic negativity. Specifically, in our approach this 
requires the calculation of the hydrodynamic behaviour of the moments of \emph{ad hoc} modified fermionic correlation 
functions, which are obtained in section~\ref{sec:pm} and section~\ref{sec:defect}. Our main 
result is discussed in section~\ref{sec:neg-res}. In section~\ref{sec:numerics} we benchmark our 
numerical results for the moments of the fermionic correlators (see section~\ref{sec:mom-num}) and 
the logarithmic negativity (see section~\ref{sec:num-neg}). We conclude in section~\ref{sec:concl}. 
In Appendix~\ref{sec:useful} we derive the formula for the negativity in terms of the fermionic 
correlation matrix for systems with fermion-number conservation. 

%
\section{Quantum quench in the open tight-binding chain}
\label{sec:model}

We consider the fermionic chain defined by the Hamiltonian 
\begin{equation}
	\label{eq:ham}
	H=\frac{1}{2}\sum_{j=1}^L(c^\dagger_j c_{j+1}+c^\dagger_{j+1}c_j). 
\end{equation}
Here $c_j^\dagger$ and $c_j$ are canonical fermionic creation and annihilation 
operators with anticommutation relations $\{c^\dagger_j,c_l\}=\delta_{jl}$ and $\{c_j,c_l\}=0$.  
For simplicity in~\eqref{eq:ham} we assume that $L$ is an even integer. 
The Hamiltonian~\eqref{eq:ham}  is diagonalized by going to Fourier space by defining the fermionic 
operators $b_k:=1/\sqrt{L}\sum_j e^{ik j} c_j$, with the quasimomentum $k=2\pi p/L$ and $p=0,1,\dots,L-1$. 
The Hamiltonian~\eqref{eq:ham} becomes diagonal as 
\begin{equation}
	\label{eq:ham-diag}
	H=\sum_k \varepsilon(k) b_k^\dagger b_k,\quad\mathrm{with}\,\,\varepsilon(k):=\cos(k), 
\end{equation}
where we defined the single-particle energy dispersion $\varepsilon(k)$. It is also convenient 
to define the group velocity of the fermionic excitations 
\begin{equation}
\label{eq:g-v}
v(k):=\varepsilon'(k)=d\varepsilon(k)/dk. 
\end{equation}
Here we focus on the nonequilibrium dynamics after the 
quench from the fermionic N\'eel state $|\mathrm{N}\rangle$ defined as 
\begin{equation}
	\label{eq:neel}
	|\mathrm{N}\rangle:=	\prod_{j=1}^{L/2}c^\dagger_{2j}|0\rangle. 
\end{equation}
	Specifically, at $t=0$ the system is prepared in $|\mathrm{N}\rangle$. 
	At $t>0$ the chain undergoes unitary dynamics under the Hamiltonian~\eqref{eq:ham}. 
The fermionic correlation function $\widetilde{C}_{jl}$ is the central object to address 
entanglement related quantities in free-fermion systems~\cite{peschel2009reduced}. 
This is defined as 
\begin{equation}
	\label{eq:c-corr}
	\widetilde{C}_{jl}:=\langle c^\dagger_j c_l\rangle=\langle\Psi(t)|c^\dagger_j c_l|\Psi(t)\rangle.  
\end{equation}
Under the closed-system dynamics implemented by $H$, the time-dependent correlation function 
$\widetilde{C}_{jl}(t)$ [cf.~\eqref{eq:c-corr}] after the N\'eel 
quench is straightforwardly obtained as 
\begin{equation}
	\label{eq:neel-c}
	\widetilde{C}_{jl}=\frac{1}{2}\delta_{jl}+\frac{1}{2}(-1)^l\int_{-\pi}^\pi\frac{dk}{2\pi}e^{ik(j-l)+2it\varepsilon(k)}. 
\end{equation}
In the following we consider the thermodynamic limit $L\to\infty$, as it is clear by the 
integration over the quasimomentum $k$. 
It is useful to exploit the invariance of the N\'eel state  under translation by 
two sites. Thus, we rewrite~\eqref{eq:neel-c} as 
\begin{equation}
	\label{eq:neel-c-1}
\left(
\begin{array}{cc}
	\widetilde C_{2j,2l} &\widetilde C_{2j,2l-1}\\
	\widetilde C_{2j-1,2l} & \widetilde C_{2j-1,2l-1}
\end{array}
\right)=\int_{-\pi}^\pi\frac{dk}{2\pi}e^{2ik(j-l)} \hat t(k), \quad\mathrm{with}\,\, j,l\in[1,L/2]. 
\end{equation}
The factor $2$ in the exponent in the integral reflects translation 
invariance by two sites. In~\eqref{eq:neel-c-1} 
we have introduced the $2\times 2$ matrix $\hat t(k)$ as 
\begin{equation}
	\label{eq:tk}
\hat t(k)=\frac{1}{2}\left(
\begin{array}{cc}
	1+e^{2it\varepsilon(k)} & -e^{2it\varepsilon(k)-ik}\\
	e^{2it\varepsilon(k)+ik} & 1-e^{2it\varepsilon(k)}
\end{array}
\right). 
\end{equation}
We can conveniently rewrite $\hat t(k)$ in terms of Pauli matrices as 
\begin{equation}
	\hat t(k)=\frac{1}{2}\Big[\mathds{1}_2+\sigma_{-i}^{(k)}e^{2it\varepsilon(k)}\Big], \quad\mathrm{with}\,\,
	\sigma_{\pm i}:=\sigma_z\pm i\sigma_y, 
	\label{eq:tk_second}
\end{equation}
where we introduce the rotated Pauli matrices $\sigma_\alpha^{(k)}$ as 
\begin{equation}
	\label{eq:sigma-rot}
	\sigma_\alpha^{(k)}:=e^{-ik/2\sigma_z}\sigma_\alpha e^{ik/2\sigma_z}, \quad\alpha=x,y,z
\end{equation}
with  $\sigma_{\alpha}$ the standard Pauli matrices.

\subsection{Lindblad evolution in the presence of gain and loss  dissipation}
\label{sec:lin}

In this work we study the out-of-equilibrium  dynamics in the tight-binding 
chain (cf.~\eqref{eq:ham})  with fermionic gain and loss processes. We employ the 
formalism of quantum master equations~\cite{petruccione2002the}. 
The Lindblad equation describes the evolution of the density matrix $\rho_t$ 
of the full system  as  
\begin{equation}
\label{eq:lind}
\frac{d\rho_t}{dt}=\mathcal{L}(\rho_t):=-i[H,\rho_t]+\sum_{j=1}^L\sum_{\alpha=\pm}\left(L_{j,\alpha}\rho_t 
L_{j,\alpha}^\dagger-\frac{1}{2}\left\{L_{j,\alpha}^\dagger L_{j,\alpha},\rho_t\right\}\right)\, .
\end{equation}
Here, $L_{j,\alpha}$ are the so-called Lindblad 
jump operators, which  are defined as  
$L_{j,-}:=\sqrt{\gamma^-}c_j$ and $L_{j,+}:=\sqrt{\gamma^+}c_j^\dagger$, 
with $\gamma^\pm$ the gain and loss rates.  
Eq.~\eqref{eq:lind} describes single-site incoherent absorption and emission 
of fermions which are homogeneous along the chain. 

For free-fermion systems it is straightforward to obtain from~\eqref{eq:lind} 
an equation for the fermionic two-point function 
$C_{jl}=\langle c^\dagger_j c_l\rangle=\mathrm{Tr}(c^\dagger_j c_l\rho_t)$. 
The time-evolved matrix $C(t)$ is given by~\cite{carollo2022dissipative} 
\begin{equation}
C(t)=e^{t \Lambda}C(0)e^{t \Lambda^\dagger} +\int_0^tdz 
\, e^{(t-z)\, \Lambda}\Gamma^+e^{(t-z)\, \Lambda^\dagger}\, .
\label{eq:CovMat}
\end{equation}
Here $\Lambda=i h-1/2(\Gamma^++\Gamma^-)$, where $h$ encodes the effects of the Hamiltonian. 
Eq.~\eqref{eq:CovMat} is obtained from~\eqref{eq:lind} by using that 
$d C_{jl}(t)/dt=\mathrm{Tr}(a^\dagger_j a_l\rho_t)$, with ${\mathcal L}$ defined 
in~\eqref{eq:lind}, and by applying Wick theorem (see Ref.~\cite{landi2022nonequilibrium} for further details). 
For the tight-binding chain considered here in~\eqref{eq:ham} 
we have $h_{jl}=1/2(\delta_{j+1,l}+\delta_{j,l+1})$.  
In~\eqref{eq:CovMat}, $\Gamma^\pm$ are $L\times L$ matrices describing gain 
and loss processes.  Here we have $\Gamma^\pm_{jl}=\gamma^\pm\delta_{jl}$, reflecting 
that gain/loss dissipation acts separately on the different sites. 
The diagonal structure of the matrices $\Gamma^\pm_{jl}$ implies that  $C_{jl}$ can be rewritten in 
terms of the correlation matrix $\widetilde C_{jl}(t)$ describing the quench 
in the absence of dissipation, i.e., with $\gamma^\pm=0$. 
Precisely, one has    
\begin{equation}
	\label{eq:C-map}
	C=n_\infty(1-b)\mathds{1}+b \widetilde C, 
	\quad b:=e^{-(\gamma^++\gamma^-)t},\,n_\infty:=\frac{\gamma^+}{\gamma^++\gamma^-}. 
\end{equation}
Moreover, Eq.~\eqref{eq:C-map} suggests that it is convenient 
to modify the matrix $\hat t(k)$ (cf.~\eqref{eq:tk}), introducing 
$\hat t'(k)$ as 
\begin{multline}
	\hat t'(k)=n_{\infty}(1-b)\mathds{1}_2+b \hat t(k)=\\
	\frac{1}{2}\left(
	\begin{array}{cc}
		2n_\infty(1-b)+b +b e^{2it\varepsilon(k)} & -be^{2it\varepsilon(k)-ik}\\
		be^{2it\varepsilon(k)+ik} & 	2n_\infty(1-b)+b -b e^{2it\varepsilon(k)} 
	\end{array}
	\right)
\end{multline}
which can also be written as 
\begin{equation}
	\label{eq:t-prime}
	\hat t'(k)=\frac{1}{2}\Big[a\mathds{1}_2+b\sigma_{-i}^{(k)}e^{2it\varepsilon(k)}\Big], \quad\mathrm{with}\,\, 
	a:=2n_\infty(1-b)+b, 
\end{equation}
where $\sigma_{-i}^{(k)}$ are defined in~\eqref{eq:tk_second}-\eqref{eq:sigma-rot}.

\subsection{Quasiparticle picture for free systems with quadratic dissipation}
\label{sec:quasi-diss}

Our goal is to determine the dynamics of the logarithmic negativity after the fermionic 
N\'eel quench in the tight-binding chain with gain and loss dissipation. Here  
we review the quasiparticle picture for 
free-fermion and free-boson systems in the presence of quadratic dissipation~\cite{carollo2022dissipative}. 
The quasiparticle picture for the entanglement 
dynamics~\cite{calabrese2005evolution,fagotti2008evolution,alba2017entanglement,alba2021generalized} can be 
generalized to describe the dynamics of quantum entropies, such as the von Neumann entropy and 
the R\'enyi entropies, and the mutual information~\cite{alba2021spreading,carollo2022dissipative,alba2022hydrodynamics} 
in the presence of generic \emph{quadratic} dissipation~\cite{prosen2008third}. 
Let us consider the R\'enyi entropies $S_A^{\scriptscriptstyle(n)}$ of a 
subsystem $A$ of length $\ell$ embedded in an infinite chain (see Fig.~\ref{fig0:bip}). 
The R\'enyi entropy of $A$ is given as~\cite{carollo2022dissipative} 
\begin{equation}
	\label{eq:ent-quasi}
	S_A^{(n)}(t)=\ell\int_{-\pi}^\pi\frac{dk}{2\pi}\Big[s_k^{(n),\mathrm{YY}}(t)-s_k^{(n),\mathrm{mix}}(t)\Big]
	\min(1,2|v(k)|t/\ell)+\ell\int_{-\pi}^\pi\frac{dk}{2\pi}s_k^{(n),\mathrm{mix}}(t), 
\end{equation}
where $v(k)$ (cf.~\eqref{eq:ham-diag} for the result in the tight-binding chain) is the fermion group velocity, 
which depends on the dispersion relation of the model. 
Crucially, Eq.~\eqref{eq:ent-quasi} holds in the standard hydrodynamic limit $t,\ell\to\infty$ with their ratio 
fixed, which is the regime of validity for the 
standard 
quasiparticle picture for the entanglement spreading after quantum 
quenches~\cite{fagotti2008evolution,alba2017entanglement,alba2018entanglement}. In the 
presence of quadratic dissipation one has to 
take the weak-dissipation limit $\gamma\to0$, with $\gamma\ell$ fixed, to ensure a nontrivial dynamics. 
Here $\gamma$ is the relevant dissipation rate, which measures the strength of the dissipative processes. 
For gain/loss dissipation this is the rate $\gamma^\pm$ (cf.~\eqref{eq:lind}). 
The reason for taking the weakly-dissipative hydrodynamic limit is that 
at finite dissipation rate and for most dissipators, in the limit $t,\ell\to\infty$ and fixed $t/\ell$ 
one obtains a trivial scaling behaviour because the entropies and the mutual information would converge 
immediately to their stationary value, which for the
mutual information is zero. 
In some cases, 
it is possible to apply Eq.~\eqref{eq:ent-quasi} away from the weak-dissipation limit 
after redefining the group velocities $v(k)$ of the quasiparticles~ and rescaling by an exponential 
factor the entropies~\cite{starchl2022relaxation}. 

Let us now discuss the structure of~\eqref{eq:ent-quasi}. 
The first term has a similar structure as in the case without 
dissipation~\cite{fagotti2008evolution,alba2017entanglement}. In the absence of dissipation 
$s_k^{\scriptscriptstyle (n),\mathrm{mix}}=0$ and $s_k^{\scriptscriptstyle(n),\mathrm{YY}}$ do not 
depend on time. Thus, Eq.~\eqref{eq:ent-quasi} describes a linear growth of the entropies up to $t<\ell/(2v_\mathrm{max})$, 
with $v_\mathrm{max}$ the maximum velocity. At asymptotically long times $t\to\infty$ 
$S_A^{\scriptscriptstyle(n)}$ saturates to a volume-law behavior. Eq.~\eqref{eq:ent-quasi}  
admits an interpretation in terms of entangled quasiparticle pairs~\cite{calabrese2005evolution}. 
After the quench, pairs of entangled quasiparticles are created uniformily in the systems. 
The quasiparticles travel as free particles. 
At the generic time $t$ the entanglement between $A$ and the rest is proportional to the number 
of shared entangled pairs, i.e., pairs that have one quasiparticle in $A$ and the 
other one in the complement of $A$. The entanglement content of the quasiparticles, 
i.e., their contribution to $S_A^{\scriptscriptstyle(n)}$ is given by the 
Yang-Yang entropies~\cite{alba2018entanglement} $s_k^{\scriptscriptstyle(n),\mathrm{YY}}$. 

The Yang-Yang entropies are determined by the density  $\rho_k$ of the Bogoliubov 
modes $b_k$ (cf.~\eqref{eq:ham-diag}) 
that diagonalize the model. The density is 
calculated over the pre-quench initial state $|\psi_0\rangle$. Specifically, we have 
\begin{equation}
\label{eq:YY-def}
	s_k^{(n),\mathrm{YY}}:=\frac{1}{1-n}\int_{-\pi}^\pi\frac{dk}{2\pi}\ln\Big(\rho_k^n+(1-\rho_k)^n\Big), 
	\quad\mathrm{with}\,\,\rho_k:=\langle\psi_0|b^\dagger_kb_k|\psi_0\rangle. 
\end{equation}
Here $s_k^{\scriptscriptstyle(n),\mathrm{YY}}$ is the density of R\'enyi entropy of the 
Generalized Gibbs Ensemble~\cite{polkovnikov2011colloquium,calabrese2016introduction,essler2016quench,vidmar2016generalized,caux2013time,caux2016the} 
(GGE) that describes local properties of the steady state after the quench.

This scenario changes dramatically in the presence of quadratic dissipation~\cite{carollo2022dissipative}. 
First, the new term $s_k^{\scriptscriptstyle(n),\mathrm{mix}}$ appears. This is 
purely dissipative and it can be obtained as the density of R\'enyi entropies of the full system. 
Indeed, the first term in~\eqref{eq:ent-quasi} cannot contribute to 
the entropies of the full system because it describes the contribution of correlated pairs that 
are shared between $A$ and its complement. If $A$ is the full system both members of a correlated pair 
are within $A$ and hence they cannot contribute. This means that only the second term in~\eqref{eq:ent-quasi} 
contributes to the full-system entropy. 
Clearly, in the unitary case the full system is in a pure state at any time and 
$s_k^{\scriptscriptstyle(n),\mathrm{mix}}=0$. For free-fermion and free-boson models $s_k^{\scriptscriptstyle(n),\mathrm{mix}}$ 
is straightforwardly extracted from the two-point correlation function in momentum space~\cite{carollo2022dissipative,alba2022hydrodynamics}. 
As it is clear from~\eqref{eq:ent-quasi}, dissipation  affects the correlation between the  quasiparticles 
as well. First, the same $s_k^{\scriptscriptstyle (n),\mathrm{mix}}$ appears in the first term in~\eqref{eq:ent-quasi}. 
The minus sign reflects that dissipation diminishes the correlation of the pairs. 
Moreover,  although $s_k^{\scriptscriptstyle (n),\mathrm{YY}}$ 
in~\eqref{eq:ent-quasi} has the form of a  Yang-Yang entropy (cf.~\eqref{eq:YY-def}), the density $\rho_k$ from which 
it is obtained is no longer 
that of the original modes $b_k$. It has been conjectured in Ref.~\cite{carollo2022dissipative} that 
in the presence of dissipation the density $\rho_k$ to be used in~\eqref{eq:YY-def} is that of the  
eigenmodes $\beta_k$ of the map $\mathcal{L}^*$, which is the dual map --- the one acting on observables --- 
of the generator $\mathcal{L}$ appearing in Eq.~\eqref{eq:lind}. 
In the weak-dissipation limit $\beta_k$ generically satisfy~\cite{carollo2022dissipative}  
\begin{equation}
	\label{eq:eig-action}
	{\cal L}^*(\beta_k)=-\left(\frac{\gamma_k}{2}+i\varepsilon(k)\right)\beta_k, 
\end{equation}
where $\varepsilon(k)$ is the dispersion of the model without dissipation, and $\gamma_k$ are dissipation 
rates that are easily calculable. Both $\varepsilon(k)$ and $\gamma_k$ are real. 
 For the case of free fermion with gain and losses,  
Eq.~\eqref{eq:eig-action} simplifies because $\beta_k$ coincides with the Fourier transformed 
fermionic operators $c_k$. Indeed, one can easily check that Eq.~\eqref{eq:eig-action} is satisfied 
with $\gamma_k=\gamma^++\gamma^-$ and $\varepsilon_k=\cos(k)$. The first term originates from the 
dissipative part (cf. second term in~\eqref{eq:lind}), whereas the second one from the unitary part of 
the Liouvillian (first term in in~\eqref{eq:lind}). Notice that for gain and loss dissipation 
Eq.~\eqref{eq:eig-action} holds also at generic $\gamma^\pm$, i.e., away from the weak-dissipation 
limit. However, Eq.~\eqref{eq:eig-action} is expected to be valid in general in the weak-dissipation 
limit because the eigenvectors and eigenvalues of ${\mathcal L}^*$ should be linear in the dissipation rates, 
in the limit of weak dissipation. Moreover, in the absence of dissipation, $\beta_k$  become 
the fermionic operators that diagonalize the system, and the eigenvalue of ${\mathcal L}^*$ becomes 
$-i\varepsilon(k)$.

Now,   
$s_k^{\scriptscriptstyle(n),\mathrm{YY}}$ in~\eqref{eq:ent-quasi} is the Yang-Yang entropy~\eqref{eq:YY-def} calculated from the density 
$\rho_k$ of the modes $\beta_k$, i.e., $\rho_k=\langle\psi_0|\beta_k^\dagger\beta_k|\psi_0\rangle$. 
In contrast with the density of the modes $b_k$, which in the unitary case is time-independent, the density of 
$\beta_k$ is time-dependent, implying that $s_k^{\scriptscriptstyle(n),\mathrm{YY}}$ depends on 
time. By computing $\mathcal{L}^*(\beta_k^\dagger \beta_k)$ and using also \eqref{eq:eig-action}, the evolution of 
$\rho_k=\langle\psi_0|\beta_k^\dagger\beta_k|\psi_0\rangle$ in the weakly-dissipative 
limit is obtained as~\cite{carollo2022dissipative}  
\begin{equation}
	\label{eq:rho-t}
	\rho_k(t)=e^{-\gamma_k t}\rho_k(0)+\frac{\alpha_k}{\gamma_k}(1-e^{-\gamma_k t}). 
\end{equation}
Here $\alpha_k$ is, again, a function that depends on the dissipation and 
that can be easily calculated
for generic free systems with quadratic dissiption~\cite{carollo2022dissipative,alba2022hydrodynamics}. 
Eq.~\eqref{eq:ent-quasi} was derived \emph{ab initio} for 
a quench in the Kitaev chain with arbitrary quadratic dissipation in Ref.~\cite{alba2022hydrodynamics}. 

For the case of diagonal gain and loss dissipation, the decay rates $\gamma_k$ in~\eqref{eq:eig-action} 
do not depend on $k$, and one has that $\gamma_k=\gamma^++\gamma^-$, 
and $\alpha_k=\gamma^+$ (cf.~\eqref{eq:rho-t}). It is also important to stress that 
while for generic dissipation the modes $\beta_k$ are different from the original Bogoliubov 
modes $b_k$ (see, for instance, 
Ref.~\cite{alba2022hydrodynamics}), for 
several types of dissipation one has that $\beta_k=b_k$. 
The gain/loss dissipation that we treat here provides one of 
the simplest examples for which this happens.

\section{Fermionic logarithmic negativity}
\label{sec:negativity}

Here we are interested in the entanglement between two non complementary 
regions $A_1$ and $A_2$ (see Fig.~\ref{fig0:bip}). 
Let us  first introduce the  
R\'enyi entropies $S_W^{(n)}$ of a subsystem $W=A_1,A_2,A_1\cup A_2$, 
which are defined as 
\begin{equation}
	\label{eq:renyi-def}
	S_W^{(n)}:=\frac{1}{1-n}\ln\Big(\mathrm{Tr}\rho_W^n\Big). 
\end{equation}
Here we often consider the von Neumann entropy, which corresponds to the 
limit $n\to1$~\cite{amico2008entanglement,calabrese2009entanglement,eisert2010colloquium,laflorencie2016quantum}. 

Crucially, since the interval $A=A_1\cup A_2$ in Fig.~\ref{fig0:bip} is in 
general in a mixed state, neither the von Neumann nor 
the R\'enyi entropies can be used to quantify the entanglement between 
$A_1$ and $A_2$~\cite{amico2008entanglement,calabrese2009entanglement,eisert2010colloquium,laflorencie2016quantum}. 
Instead, one can use the logarithmic negativity~\cite{lee2000partial,vidal2002computable,eisert2006entanglement,plenio2005logarithmic,calabrese2012entanglement} 
${\cal E}$, which is a computable entanglement measure for mixed states. 
To define ${\cal E}$ one has to introduce  the partially-transposed 
density matrix $\rho^{T_2}_A$. The partial transposition is taken  
with respect to one of the intervals (here $A_2$). 
$\rho^{T_2}_A$ is defined as 
\begin{equation}
	\langle e^{(1)}_i,e^{(2)}_j|\rho_A^{T_2}| e^{(1)}_k, e^{(2)}_l\rangle=
	\langle e_i^{(1)},e_l^{(2)}|\rho_A|e_k^{(1)},e_j^{(2)}\rangle, 
\end{equation}
with $e^{(1)}_i,e_j^{(2)}$ two bases for $A_1$ and $A_2$, respectively. 
The partial transpose is not positive-definite, and  its negative 
eigenvalues quantify the  entanglement between the two intervals. 
The logarithmic negativity is defined as 
\begin{equation}
	\label{eq:neg-def0}
	{\cal E}=\ln(\mathrm{Tr}|\rho^{T_2}_{A}|). 
\end{equation}
For free-boson systems the negativity can be computed from the two-point 
correlation function~\cite{audenaert2002entanglement}. 
For free-fermion systems the partially transposed reduced density matrix 
can be decomposed as~\cite{eisler2015on} 
\begin{equation}
	\label{eq:opm}
	\rho^{T_2}_A=e^{-i\pi/4} O_++e^{i\pi/4}O_-, 
\end{equation}
where $O_\pm$ are gaussian operators. Crucially, while the spectrum of 
$O_\pm$ can be effectively computed from that of the fermionic two-point 
function, this cannot be done for $\rho_A^{T_2}$. As a consequence, the negativity cannot 
be easily calculated, not even  for free-fermion models, although several results have been 
obtained in the literature~\cite{coser2015partial,coser2016towards,herzog2016estimation} for the moments of 
the partial transpose.  
Recently, it has been shown that starting from the decomposition~\eqref{eq:opm}, it is possible 
to construct an alternative measure of entanglement for mixed-state systems. This has been dubbed 
fermionic negativity~\cite{shapourian2017many,shapourian2017partial,eisert2018entanglement,shapourian2019entanglement}. 
The fermionic negativity is defined as 
\begin{equation}
	\label{eq:e-f}
	{\cal E}:=\ln\mathrm{Tr}\sqrt{O_+O_-}. 
\end{equation}
Notice that here we use the same symbol ${\cal E}$ for both the standard negativity (cf.~\eqref{eq:neg-def0}) 
and the fermionic one. In the following sections we will always refer to the fermionic negativity. 

Since the product $O_+O_-$ in~\eqref{eq:e-f} is a gaussian 
operator because $O_\pm$ is gaussian,  
the fermionic negativity~\eqref{eq:e-f} can be computed 
effectively in terms of fermionic two-point functions. 
The central object is the fermionic correlation matrix $C_{jl}$ (cf.~\eqref{eq:CovMat}).  
Let us define the matrix $G_{jl}$ as 
\begin{equation}
	\label{eq:g-def}
	G_{jl}:=2C_{jl}-\delta_{jl}. 
\end{equation}
We now consider the partition in Fig.~\ref{fig0:bip}. 
We focus on two intervals $A_1$ and $A_2$ of equal length $\ell$ at distance $d$. 
We now define $G^{\alpha\beta}_{jl}$ with $\alpha,\beta=1,2$ as the restricted correlator with 
$j\in A_\alpha$ and $l\in A_\beta$. The matrix $G_A$ is rewritten as 
\begin{equation}
	\label{eq:gij}
G_A=\left(
\begin{array}{cc}
	G^{11} & G^{12}\\
	G^{21} & G^{22}
\end{array}
\right),
\end{equation}
where $G^{\alpha\beta}$ are $\ell\times\ell$ matrices. 
We now define the matrices $G_A^\pm$ as 
\begin{equation}
	\label{eq:gpm}
	G_A^\pm=\left(
\begin{array}{cc}
	G^{11} & \pm i G^{12}\\
	\pm i G^{21} & -G^{22}
\end{array}
\right)
\end{equation}
 These are the covariance matrices of the operators $O^\pm$ 
 introduced in~\eqref{eq:opm}.  
Finally, the negativity is a function of the spectrum of 
$C_A$, which the fermionic correlator $C$ restricted to $A$,  
and that of $G_A^{\mathrm{T}}$, which is defined as 
\begin{equation}
	\label{eq:gt}
	G_A^{\mathrm{T}}:=\frac{1}{2}\Big[\mathds{1}_{2\ell}-(\mathds{1}_{2\ell}+G^+_AG^-_A)^{-1}(G^+_A+G^-_A)\Big]. 
\end{equation}
Here $\mathds{1}_{2\ell}$ is the $2\ell\times 2\ell$ identity matrix. 
$G_A^{\mathrm{T}}$ is the covariance matrix of the product $O^+ O^-/\mathrm{Tr}(O^+O^-)$. 
The negativity is defined as~\cite{shapourian2019finite} 
\begin{equation}
	\label{eq:neg-comput}
	\mathcal{E}:=\sum_{j=1}^{2\ell}\ln[\mu_j^{1/2}+(1-\mu_j)^{1/2}]+
	\sum_{j=1}^{2\ell}\frac{1}{2}\ln[\lambda_j^2+(1-\lambda_j)^2]. 
\end{equation}
where $\mu_j$ are the eigenvalues of $G_A^{\mathrm{T}}$ and $\lambda_j$ of $C_A$. 
The second term in~\eqref{eq:neg-comput} originates from the normalization $\mathrm{Tr}(O^+O^-)=
\mathrm{Tr}\rho_A^2$ (see, for instance, Ref.~\cite{eisert2016entanglement}). 
Importantly, Eq.~\eqref{eq:neg-comput} holds for free-fermion systems 
that preserve the fermion 
number. For generic free-fermion systems  a generalization of~\eqref{eq:neg-comput} 
exists in terms of the correlation function of Majorana fermions~\cite{shapourian2019entanglement,eisert2018entanglement}. 
In the presence of gain/loss dissipation the fermion number is not preserved, although at  
any time one has $\langle c^\dagger_j c^\dagger_l\rangle=\langle c_j c_l\rangle=0$. In Appendix~\ref{sec:useful} 
we show that this condition is sufficient to ensure the validity of~\eqref{eq:neg-comput}. 

The logarithmic negativity has been employed to characterize entanglement in systems of harmonic 
oscillators~\cite{audenaert2002entanglement,ferraro2008thermal,anders2008entanglement,marcovitch2009critical,eisler2014entanglement,sherman2016nonzero,denobili2016entanglement}, 
spin models~\cite{alba2013entanglement,chung2014entanglement,calabrese2013entanglement,wu2020entanglement,lu2020entanglement,wichterich2009scaling,wichterich2010universality,bayat2012entanglement,ruggiero2016entanglement,turkeshi2020entanglement,mbeng2017negativity,wald2020entanglement,shapourian2021entanglement}, Conformal Field Theory (CFT)~\cite{calabrese2012entanglement,calabrese2013entanglementnegativity,calabrese2014finite,wen2015entanglement,ruggiero2016negativity,alba2017entanglementspectrum,cornfeld2019measuring,kudler-flam2020the}. 
The out-of-equilibrium dynamics after quantum quenches has received a lot of 
attention~\cite{coser2014entanglement,alba2019quantum,feldman2019dynamics,kudler2020correlation,kudler2021the,gruber2020time,elben2020mixed,murciano2021quench,parez2022dynamics}. In particular, it has been shown in Ref.~\cite{alba2019quantum} that for large intervals, long times, and large distance (see Fig.~\ref{fig0:bip}) 
$\ell,t,d\to\infty$ with the ratios $\ell/t$ and $d/t$ fixed, the standard negativity 
and the fermionic one become 
\begin{equation}
	\label{eq:e-mi}
{\cal E}=\frac{1}{2}I_{A_1:A_2}^{(1/2)}, 
\end{equation}
where $I^{\scriptscriptstyle(1/2)}_{A_1:A_2}$ is the R\'enyi mutual information with 
R\'enyi index $1/2$. 
Eq.~\eqref{eq:e-mi} was proposed in Ref.~\cite{alba2019quantum}, and it was verified  for quantum quenches in 
free-boson and free-fermion systems. It is natural to expect that Eq.~\eqref{eq:e-mi} holds 
for generic interacting integrable systems. Very recently, it has been shown that in the early-time regime 
Eq.~\eqref{eq:e-mi} holds true in quenches with generic local quantum circuits~\cite{bertini2022entanglement}. 
Eq.~\eqref{eq:e-mi} is intriguing because in general the mutual information between  
$A_1$ and $A_2$ is not a good measure of their entanglement but only 
an upper bound~\cite{wolf2008area}.

\section{Warm-up: quantum entropies in the presence of gain and loss dissipation}
\label{sec:warm-up}

As a warm-up, before deriving the quasiparticle picture for the fermionic negativity, here we 
provide an alternative direct derivation of the results of Ref.~\cite{alba2021spreading}.
In contrast with Ref.~\cite{alba2021spreading} the derivation that we present here is 
\emph{ab initio}, although we rely on the same analytic techniques employed in Ref.~\cite{calabrese2012quantum}. 
In section~\ref{sec:moments} we derive the behavior 
of the moments of the correlation matrix $\mathrm{Tr} (C_A^n)$. In section~\ref{sec:renyi} we 
discuss the R\'enyi entropies. 

\subsection{Moments of the correlators $\mathrm{Tr}(C^n_A)$} 
\label{sec:moments}

Here we determine the scaling of $\mathrm{Tr}(C_A^n)$ in the hydrodynamic limit $t,\ell\to\infty$, 
with their ratio  $t/\ell$ fixed. At the end of the derivation we will also discuss the 
weakly-dissipative limit by taking $\gamma^\pm\to0$ with the product $\gamma^\pm\ell$ 
fixed. To derive our main results  we employ the approach of Ref.~\cite{calabrese2012quantum}.
The correlator $C_A$ is defined in~\eqref{eq:C-map}. 
We can use the trivial identity 
\begin{equation}
	\label{eq:id}
	\sum_{z=1}^{\ell/2} e^{2izk}=\frac{\ell}{4}
	\int_{-1}^{1}d\xi w([k]_\pi)e^{i(\ell\xi+\ell+2)[k]_\pi/2},\quad\mathrm{with}\, w(k):=\frac{k}{\sin(k)}, 
\end{equation}
where we introduced the notation $[x]_\pi=x\,\mathrm{mod}\,\pi$. The $\mathrm{mod}\,\pi$ 
reflects the factor $2$ in the exponent in the left hand side in~\eqref{eq:id}, and 
it is due to the fact that the initial state is not invariant under one-site translation, although 
it is invariant under two-site translations. 
Notice that~\eqref{eq:id} is different from a similar identity used in 
Ref.~\cite{calabrese2012quantum}, which deals with one-site translation invariant initial states. 
Eq.~\eqref{eq:id} allows us to write 
\begin{equation}
	\label{eq:tr-0}
	\mathrm{Tr}(C_A^n)=\Big(\frac{\ell}{4}\Big)^{n}\int_{[-\pi,\pi]^n}\frac{d^nk}{(2\pi)^n}\int_{-1}^1 d^n\xi D(\{k\})F(\{k\})
	e^{i\ell\sum_{j=0}^{n-1}\xi_{j+1}([k_{j+1}-k_j]_\pi)/2}. 
\end{equation}
Here we defined 
\begin{align}
	\label{eq:D-def}
	D(\{k\})&=\prod_{j=0}^{n-1}w([k_j-k_{j-1}]_\pi)\\
	\label{eq:F-def}
	F(\{k\})&=\mathrm{Tr}\prod_{j=0}^{n-1}\hat t'(k_j), 
\end{align}
where $t'(k)$ is defined in~\eqref{eq:t-prime}. 
In deriving~\eqref{eq:tr-0} from~\eqref{eq:id}, 
we neglect the factor $\ell+2$  because it contributes with a  phase. 
Notice that the quasimomenta in $F$ (cf.~\eqref{eq:F-def}) are not defined $\mathrm{mod}$ $\pi$. 
It is convenient to define new variables $\zeta_j$ as
\begin{align}
&\zeta_0=\xi_1\\
&\zeta_i=\xi_{i+1}-\xi_i,\quad i\in[1,n-1]. 
\end{align}
This allows us to write~\eqref{eq:tr-0} as 
\begin{equation}
	\label{eq:tr}
	\mathrm{Tr}(C_A^{n})=
	\Big(\frac{\ell}{4}\Big)^{n}\int\limits_{[-\pi,\pi]^{n}}\frac{d^{n}k}{(2\pi)^{n}}
	\int_{R_\xi} d^{n}\zeta_i D(\{k\})F(\{k\})e^{-i\ell\sum_{j=0}^{n-1}\sum_{l=0}^j\zeta_l([k_{j+1}-k_{j}]_\pi)/2}. 
\end{equation}
Here the integration domain $R_\xi$ for $\zeta_i$ is 
\begin{equation}
	\label{eq:domain}
	R_\xi:-1\le\sum_{j=0}^{p-1}\zeta_j\le1, \quad p\in[1,n]. 
\end{equation}
The strategy to determine the behaviour of~\eqref{eq:tr}  in the space-time scaling limit is to 
use the stationary phase approximation for the integrals over $k_1,\dots,k_{n-1}$ and $\zeta_1,\dots,\zeta_{n-1}$. 
It is easy to check that stationarity with respect to the variables $\zeta_1,\dots,\zeta_{n-1}$ implies that 
\begin{equation}
	\label{eq:k-stat}
	[k_{j+1}-k_{j}]_\pi=0,\quad \forall j\in[0,n-1]. 
\end{equation}
This also implies that the integrand in~\eqref{eq:tr} does not depend on $\zeta_0$. 
Thus, the integration over $\zeta_0$ is trivial and we obtain 
\begin{equation}
	\label{eq:tr-1}
	\mathrm{Tr}(C^{n}_A)=\Big(\frac{\ell}{4}\Big)^{n}\int\limits_{[-\pi,\pi]^{n}}\frac{d^{n}k}{(2\pi)^{n}}
	\int d^{n-1}\zeta_i D(\{k\})F(\{k\})e^{-i\ell\sum_{j=0}^{n-1}\sum_{l=1}^j\zeta_l([k_{j+1}-k_{j}]_\pi)/2}\mu(\{\zeta_j\}), 
\end{equation}
where we introduced the integration measure $\mu$ as 
\begin{equation}
	\label{eq:measure}
	\mu(\{\zeta_j\})=\max\Big[0,\min\limits_{j\in[0,n-1]}
	\Big[1-\sum_{k=1}^j\zeta_k\Big]+\min\limits_{j\in[0,n-1]}\Big[1+\sum_{k=1}^j\zeta_k\Big]\Big]. 
\end{equation}

Now we can replace $k_j\to k_0$ in $D(\{k_j\})$ because it depends only on 
$[k_j-k_{j-1}]_\pi$  to obtain 
\begin{align}
	\label{eq:D-1}
	& D(\{k_j\})\to 1. 
\end{align}
The  function $F(\{k_j\})$ requires some care. 
We can expand it as 
\begin{equation}
	\label{eq:inte-d}
	F=
	\frac{1}{2^{n}}\sum_{p=0}^{n-1}\sum_{j_1<\dots<j_p=1}^{n-1}
	\mathrm{Tr}[(a\mathds{1}_2+b\sigma_{-i}^{(k_0)}e^{2it\varepsilon(k_0)})b^p\sigma_{-i}^{(k_{j_1})}\sigma_{-i}^{(k_{j_2})}\cdots \sigma_{-i}^{(k_{j_p})}]e^{2it\sum_{l=1}^p\varepsilon(k_{j_l})}, 
\end{equation}
where $a,b$ and $\sigma_{-i}$ are defined in~\eqref{eq:t-prime}, and we isolated the term with $k_0$. 
It is worth noticing that in Eq.~\eqref{eq:inte-d} the time-dependent term appearing in the exponent 
	is a number. This is specific of the tight-binding chain. For generic quenches in the $XY$ chain the term in 
	the exponent is a matrix~\cite{calabrese2012quantum}. 

To perform the trace in~\eqref{eq:inte-d} we observe that $\sigma_{-i}^{\scriptscriptstyle(k)}\sigma_{-i}^{\scriptscriptstyle(k')}=0$ if  $k\to k'$. 
On the other hand, one has that that $\mathrm{Tr}(\sigma_{-i}^{\scriptscriptstyle(k_0)}\sigma_{-i}^{\scriptscriptstyle(k_{j_1})}\dots
\sigma^{\scriptscriptstyle(k_{j_{p-1}})}_{-i})=2^{p+1}$ if from the stationary solution~\eqref{eq:k-stat} 
one selects the alternating pattern as $k_{j_1}=k_0+\pi,k_{j_2}=k_0,\dots$, and it vanishes otherwise. 
For the second term, one has  $\mathrm{Tr}(\sigma_{-i}^{\scriptscriptstyle(k_{j_1})}\sigma_{-i}^{\scriptscriptstyle(k_{j_2})}\dots
\sigma^{\scriptscriptstyle(k_{j_p})}_{-i})=2^{p+1}$ for both the alternating patterns $k_{j_1}=k_0,k_{j_2}=k_0+\pi,\dots$ and 
$k_{j_1}=k_{0}+\pi,k_{j_{2}}=k_0,\dots$. 
This implies that the first term  in the trace in~\eqref{eq:inte-d} gives nonzero contribution 
only for even $p$, whereas the second term contributes to odd $p$. 
Thus, we can rewrite~\eqref{eq:inte-d} as 
\begin{multline}
	\label{eq:inte-1-d}
F(\{k_j\})\propto 
2^{-n}\sum_{p=0}^{\lfloor (n-1)/2\rfloor }\binom{n-1}{2p}a^{n-2p} (2b)^{2p}e^{2it\sum_{l=1}^{2p}\varepsilon(k_{l})}\\+
2^{-n}\sum_{p=0}^{\lfloor (n-2)/2\rfloor}\binom{n-1}{2p+1}a^{n-2p-2}(2b)^{2p+2}e^{2it\sum_{l=0}^{2p+1}\varepsilon(k_l)}. 
\end{multline}
Here we used the invariance under relabelling of the momenta $k_{j_l}$ to replace $k_{j_l}\to k_l$ in 
the phase factor. However, we are not allowed to replace $k_l$ with their stationary values 
before performing the stationary phase approximation. The 
binomials in~\eqref{eq:inte-1-d} are  the number of terms containing $2p$ and $2p+1$ 
quasimomenta, and that are the same under exchange of  the momentum label. 
The factors $2^{2p}$ and $2^{2p+2}$ are the results of the trace operation. 
Finally, the proportionality symbol $\propto$ in~\eqref{eq:inte-1-d} is because there is 
an extra constant that originates from the total number of stationary points in~\eqref{eq:k-stat}. 
Indeed, in principle Eq.~\eqref{eq:k-stat} corresponds to $2^{n}$ stationary points. 
 This proliferation of the stationary points is due to the invariance under shift by $\pi$ (cf.~\eqref{eq:k-stat}), 
which reflects translation invariance by two sites. Again, this is different from Ref.~\cite{calabrese2012quantum}. 
However, the presence of the 
strings of $\sigma_{-i}^{\scriptscriptstyle(k_j)}$ selects one of the two patterns $k_1=k_0,k_2=k_0+\pi,\dots$ or 
$k_1=k_0+\pi,k_2=k_0,\dots$. For the first term in~\eqref{eq:inte-1-d} there is a remaining overall factor $2^{n-1-2p}$ 
that originates from the $n-1-2p$ quasimomenta that do not appear in the string of $\sigma_{-i}$. Moreover, there is 
an extra factor $2$ because both alternating patterns $\{\bar k_1,\bar k_2,\dots, \bar k_{2p}\}=\{k_0,k_0+\pi,k_0,\dots\}$ 
and $\{\bar k_1,\bar k_2,\dots,\bar k_{2p}\}=\{k_0+\pi,k_0,k_0+\pi,\dots\}$ contribute. This is different for the second term 
in~\eqref{eq:inte-1-d}. The stationary phase treatment of the quasimomenta that do not appear in the phase 
factor gives a factor $2^{n-2-2p}$. There is no extra 
factor $2$ because only the quasimomenta pattern $\{\bar k_1,\bar k_2,\bar k_3,\dots\}=\{k_0+\pi,k_0,k_0+\pi,\dots\}$ 
gives a nonzero contribution after taking the trace in~\eqref{eq:inte-d}. 
By putting together all the factors, the result is that one can drop the 
prefactors $2^{-n}, 2^{2p}$ and $2^{2p+2}$ in~\eqref{eq:inte-1-d} to obtain 
\begin{multline}
\label{eq:inte-2-d}
F=
\sum_{p=0}^{\lfloor (n-1)/2\rfloor }\binom{n-1}{2p}a^{n-2p} b^{2p}e^{2it\sum\limits_{l=1}^{2p}\varepsilon(k_{l})}+
\sum_{p=0}^{\lfloor n/2\rfloor-1}\binom{n-1}{2p+1}a^{n-2p-2}b^{2p+2}e^{2it\sum\limits_{l=0}^{2p+1}\varepsilon(k_l)}. 
\end{multline}
The strategy is now to apply the stationary phase approximation to 
the integral in the $2n-2$ variables $k_1,k_2,\dots,k_{n-1}, 
\zeta_1,\zeta_2,\dots,\zeta_{n-1}$.  For the first term in~\eqref{eq:inte-2-d} with 
the quasimomenta $k_1,k_2,\dots,k_{2p}$ appearing in the phase factor one obtains 
the stationary points $\bar k_j$ and $\bar\zeta_j$ as 
\begin{align}
	\label{eq:pattern}
	& \{\bar k_1,\bar k_2,\dots,\bar k_{2p}\}=\{k_0,k_0+\pi,\dots k_0+\pi\}\cup\{k_0+\pi,k_0,\dots k_0\} & \\
	\label{eq:zeta-statio}
	& \bar\zeta_{l}=\pm 4\frac{t}{\ell}(-1)^l\varepsilon'(k_0) & l=1,\dots,2p\\
	& \bar\zeta_j=0 & l>2p. 
\end{align}
Here we have to choose only one of the patterns in~\eqref{eq:pattern} because they give 
the same result, and this was already taken into account in~\eqref{eq:inte-2-d}. 
The sign of $\bar \zeta_j$ in~\eqref{eq:zeta-statio} is different for the two patterns 
in~\eqref{eq:pattern}.  However, this sign does not affect the final result. 
The reason is that $\bar \zeta_l$ enter only in the function $\mu(\{\zeta_j\})$ (cf.~\eqref{eq:measure}), 
which remains the same under change of the sign of $\bar\zeta_l$. 
The stationary phase treatment of the second term in~\eqref{eq:inte-2-d} is similar. The result is 
that only the second pattern in~\eqref{eq:pattern} contributes.  
We now use the formula for the stationary phase approximation~\cite{wong2001asymptotic}  
\begin{equation}
\label{eq:s-phase}
\int_{\mathcal{D}} d^N xp(\vec x)e^{i\ell q(\vec x)}
\rightarrow\Big(\frac{2\pi}{\ell}\Big)^{N/2}p(\vec x_0)|\mathrm{det}H|^{-1/2}\exp\Big[i\ell q(\vec x_0)+i\pi\frac{\sigma_A}{4}\Big].
\end{equation}
Here $p(\vec x)$ and $q(\vec x)$ are functions, $\mathcal{D}$ denotes the domain 
of integration and
$\ell$ is the large parameter. In~\eqref{eq:s-phase},
we denote by $\vec x_0$  the stationary point that is solution of 
$\vec\nabla q(\vec x_0)=0$. The Hessina matrix $H$ is 
given by $H=\partial_{x_i}\partial_{x_j} q(\vec x)$. 
The signature $\sigma$ of the Hessian is calculated as 
the difference between the number of positive and negative
eigenvalues of $H$, and in our case it is zero. Moreover, in our case 
$|\mathrm{det} H|^{-1/2}=2^{n-1}$, and the phase in~\eqref{eq:s-phase} 
vanishes. 

In using~\eqref{eq:s-phase}, we observe that the term with $p=0$ in the first sum in~\eqref{eq:inte-2-d} 
gives (cf.~\eqref{eq:measure}) $\mu(\{\zeta_j\})=2$. On the other hand, all the other 
terms give the same result as 
\begin{equation}
	\label{eq:mu-s}
	\mu(\{\zeta_j\})=2\max(0,1-2t/\ell|v(k_0)|), 
\end{equation}
with $v(k_0)$ the group velocity (cf.~\eqref{eq:g-v}). 
Finally, it is clear from~\eqref{eq:inte-2-d} that the term with $p=0$ contributes with 
$a^n$, whereas the remaining sum gives $((a+b)^n+(a-b)^n-2a^n)/2$. 
This implies that 
\begin{equation}
	\label{eq:final-0}
	\mathrm{Tr}(C^n_A)=\frac{a^n}{2^n}\ell+\frac{(a-b)^n+(a+b)^n-2a^n}{2^{n+1}}\int_{-\pi}^{\pi}
	\frac{dk}{2\pi}\max(0,\ell-2t|v(k)|), 
\end{equation}
where we replaced $k_0\to k$. This can be rewritten as 
\begin{multline}
	\label{eq:final-1}
	\mathrm{Tr}(C^n_A)=\ell\int_{-\pi}^\pi\frac{dk}{2\pi} \Big[\Big(\frac{a}{2}\Big)^n-\frac{(a+b)^n+(a-b)^n}{2^{n+1}}
	\Big] \min(1,2|v(k)|t/\ell)\\+
	\ell\int_{-\pi}^\pi\frac{dk}{2\pi} \frac{(a+b)^n+(a-b)^n}{2^{n+1}}. 
\end{multline}
Now, the first term in~\eqref{eq:final-1} admits a quasiparticle interpretation. Indeed, the function 
$\min(1,2|v(k)|t/\ell)$ is the number of pairs of entangled quasiparticles with quasimomenta 
$k$ and $-k$ that are shared between $A$ and the rest at time $t$. The second term is proportional to the 
volume of $A$ and it is not due to the pairs of quasiparticles. 

\subsection{R\'enyi entropies and von Neumann entropy}
\label{sec:renyi}

Eq.~\eqref{eq:final-1} allows us to obtain the behaviour of $\mathrm{Tr}({\mathcal F}(C_A))$, where 
${\mathcal F}(z)$ is an arbitrary function that admits a Taylor expansion in 
$z=0$. After expanding ${\mathcal F}(C_A)$ and using~\eqref{eq:final-1}, one obtains 
\begin{multline}
	\label{eq:F-final-1}
	\mathrm{Tr}(\mathcal{F}(C_A))=\ell\int_{-\pi}^\pi \frac{dk}{2\pi}\Big[\mathcal{F}(a/2)
-\frac{\mathcal{F}((a+b)/2)+\mathcal{F}((a-b)/2)}{2}
\Big]\min(1,2|v(k)|t/\ell) \\
	+\ell\int_{-\pi}^\pi \frac{dk}{2\pi}\frac{\mathcal{F}((a+b)/2)+\mathcal{F}((a-b)/2)}{2}. 
\end{multline}
The functions $\mathcal{F}_n(x)$  and $\mathcal{F}_1(x)$ that correspond to 
the R\'enyi entropies $S^{(n)}_A$ and the von Neumann entropy $S_A$ read 
\begin{align}
	\label{eq:renyi-f}
	& \mathcal{F}_n(x):=\frac{1}{1-n}\ln(x^n+(1-x)^{n})\\
	\label{eq:vn-f}
	& \mathcal{F}_1(x):=-x\ln(x)-(1-x)\ln(1-x). 
\end{align}
After using~\eqref{eq:renyi-f} and~\eqref{eq:vn-f} in~\eqref{eq:F-final-1} one recovers 
the results of Ref.~\cite{alba2021spreading}. 
Let us also discuss the non dissipative limit $\gamma^\pm\to0$. In that limit $a,b\to1$ 
(see~\eqref{eq:C-map} and~\eqref{eq:t-prime}), and one obtains  
\begin{equation}
	\label{eq:s-nodiss}
	\mathrm{Tr}(\mathcal{F}_n(C_A))=\int_{-\pi}^\pi\frac{dk}{2\pi}\left\{ \Big[\mathcal{F}_n(1/2)-
	\frac{{\mathcal F}_n(0)+{\mathcal F}_n(1)}{2}\Big]\min(\ell,2|v(k)|t)+
	\frac{\ell}{2}({\mathcal F}_n(0)+{\mathcal F}_n(1))\right\}. 
\end{equation}
Since ${\mathcal F}_n(0)={\mathcal F}_n(1)=0$ for any $n$, 
only the first term in the square brackets in~\eqref{eq:s-nodiss} survives. One has that 
${\mathcal F}_n(1/2)=\ln(2)$ for any $n$, which implies that for the N\'eel quench 
all the R\'enyi entropies are equal. Moreover, the density of R\'enyi entropy does not depend 
on $k$. Both these two features are specific for the N\'eel quench. 
Finally, we should observe that Eq.~\eqref{eq:renyi-f} holds in the limit $\ell,t\to\infty$ with 
the ratio $t/\ell$ finite. In particular, for this simple case of diagonal dissipation 
Eq.~\eqref{eq:renyi-f} holds also at finite $\gamma^\pm$. 
Still, in the limit $t\to\infty$ one has that $S_A^{(n)}\to 0$ for any $n$. To have a nontrivial 
dynamics we take the weakly-dissipative hydrodynamic limit $t,\ell\to\infty$, $\gamma^\pm\to0$, 
with $t/\ell$ and $\gamma^\pm\ell$ fixed~\cite{alba2021spreading,carollo2022dissipative,alba2022hydrodynamics}.

\section{Fermionic logarithmic negativity}
\label{sec:neg-calculation}

We are now ready to  derive the behaviour of the fermionic logarithmic negativity in the weakly-dissipative 
hydrodynamic limit. The strategy is as follows. We first focus on the moments $\mathrm{Tr}[(G^+G^-)^n]$. In 
the following section we drop the subscript $A$ in $G_A^\pm$, as it is clear that we will always consider 
the correlators restricted to subsystem $A$. 
This is presented in Section~\ref{sec:pm}. Then we 
determine the hydrodynamic behaviour of some modified moments of $G^+G^-$. 
These are obtained by expanding the $n$-th power of $G^+G^-$ and inserting an arbitrary number 
 $m$ of ``misplaced'' $G^\pm$. These insertions break the alternating pattern of $G^+G^-$, 
creating ``defects'' at places where the same operator is present on consecutive positions. 
The hydrodynamic limit of these defective moments is derived in Section~\ref{sec:defect}. 
Finally, in Section~\ref{sec:neg-res} we provide the result for the fermionic logarithmic 
negativity.

\subsection{Moments $\mathrm{Tr}[(G^+G^-)^n]$}
\label{sec:pm}

Let us now consider the moments of the product $G^+G^-$ (cf.~\eqref{eq:gpm}), i.e, 
$\mathrm{Tr}[(G^+G^-)^n]$ for arbitrary $n$. Here we focus on  a subsystem $A$ of 
length $2\ell$, which is further divided into two adjacent equal-length  intervals $A_1$ 
and $A_2$ (see Fig.~\ref{fig0:bip}). Let us start by defining the Fourier transform 
$\hat t''(k)$ of the matrix (cf.~\eqref{eq:gij}) $G=\mathds{1}-2C$ as  
\begin{equation}
	\hat t''(k)=a'\mathds{1}_{2}-b\sigma_{-i}^{(k)}e^{2it\varepsilon(k)},\quad a':=1-a. 	
\end{equation}
The matrices $G^\pm$ (cf.~\eqref{eq:gpm}) are written as 
\begin{equation}
	\label{eq:gpm-d}
	G^{\pm}=\int_{-\pi}^\pi\frac{dk}{2\pi} e^{2ik(j-l)}\sigma_{\mp}^{(k\ell)}
	\otimes[a'\mathds{1}_2
-b\sigma_{-i}^{(k)}e^{2it\varepsilon}],\quad j,l=1,\dots,\ell. 
\end{equation}
In~\eqref{eq:gpm-d} we defined $\sigma_{\pm}^{(k\ell)}$ as 
\begin{equation}
	\label{eq:rot-1}
	\sigma^{(k\ell)}_{\pm}:=\sigma^{(k\ell)}_z\pm\sigma^{(k\ell)}_y,   
\end{equation}
where $\sigma_{\alpha}^{(k\ell)}$ are the rotated Pauli matrices defined 
in~\eqref{eq:sigma-rot}. Notice that $\ell$ appears in the rotation 
angle in~\eqref{eq:rot-1}. This is important when using the stationary phase approximation. 
The tensor product with $\sigma_\mp^{\scriptscriptstyle(k\ell)}$ 
in~\eqref{eq:gpm-d} accounts for the fact that the indices $j,l$ 
are shifted by $\ell$ when considering the blocks  $G^{12}$ and $G^{21}$ (cf.~\eqref{eq:gij}). 
The structure in~\eqref{eq:gpm-d} 
is straightforwardly generalizable to quenches  from other initial states 
by changing the term in the square brackets. 
Similar to~\eqref{eq:tr-0}, one can write the moments $\mathrm{Tr}[(G^+G^-)^n]$ as 
\begin{equation}
	\label{eq:mom-1}
	\mathrm{Tr}[(G^+G^-)^{n}]=\Big(\frac{\ell}{4}\Big)^{2n}\int_{[-\pi,\pi]^{2n}}\frac{d^{2n}k}{(2\pi)^{2n}}
	\int_{-1}^1 d^{2n}\xi D(\{k\})F(\{k\}) e^{i\ell\sum_{j=0}^{2n-1}\xi_{j+1}([k_{j+1}-k_j]_\pi)/2}. 
\end{equation}
Here $D(\{k\})$ is the same as in section~\ref{sec:warm-up}, whereas $F(\{k_j\})$ 
is given by
\begin{equation}
	\label{eq:F-1}
	F(\{k_j\})=\Big(\mathrm{Tr}\prod_{j=0}^{n-1} \sigma_+^{(\ell k_{2j})}\sigma_-^{(\ell k_{2j+1})}\Big)
	\Big(\mathrm{Tr}\prod_{j=0}^{2n-1}(a'\mathds{1}_2-b\sigma_{-i}^{(k_j)}e^{2it\varepsilon})\Big). 
\end{equation}
To take the trace we use that 
\begin{equation}
	\label{eq:idpm}
	\sigma^{(\ell k_0)}_+\sigma^{(\ell k_1)}_-\cdots 
	\sigma_{+}^{(\ell k_{2n-2})}\sigma_-^{(\ell k_{2n-1})}=
	e^{-i\ell\sum_{j=0}^{2n-1}k_j}\prod_{j=0}^{2n-1}(e^{i k_j\ell}+e^{ik_{j+1}\ell})
	\left(
	\begin{array}{cc}
		e^{ik_{2n-1}\ell} & -i\\
		i e^{i(k_0+k_{2n-1})\ell} & e^{ik_0\ell}
	\end{array}
\right). 
\end{equation}
Notice that here we use periodic boundary conditions on the quasimomenta, meaning that 
$k_{2n}=k_0$. 
One can expand~\eqref{eq:idpm} to obtain 
\begin{multline}
\label{eq:tr-4}
\mathrm{Tr}\prod_{j=0}^{n-1} \sigma_+^{(\ell k_{2j})}\sigma_-^{(\ell k_{2j+1})}
=\\2+\sum_{z=1}^{n}\sum_{j_1<\dots<j_{2z}=0}^{2n-1} 
	(e^{i \ell (k_{j_1}-k_{j_2}+k_{j_3}-\dots -k_{j_{2z}})}+
	e^{-i \ell (k_{j_1}-k_{j_2}+k_{j_3}-\dots -k_{j_{2z}})}). 
\end{multline}
Note the alternating pattern in the exponents in~\eqref{eq:tr-4}. 
The evaluation of the trace of the right term in~\eqref{eq:F-1} can be performed as in~\eqref{eq:inte-1-d}. 
The stationary phase approximation with respect to the variables $\zeta_j$ gives $[k_{j+1}-k_j]_\pi=0$, 
i.e., the same as in~\eqref{eq:pattern}. 
The trace in the right term in~\eqref{eq:F-1} is the same as in~\eqref{eq:inte-2-d} after redefining $a\to a'$, 
and $n=2p$. 
One obtains 
\begin{multline}
	\label{eq:F-2}
	2^{-2n}\mathrm{Tr}\prod_{j=0}^{2n-1}\hat t''(k_j)=
\sum_{j=0}^{\lfloor (2n-1)/2\rfloor }\binom{2n-1}{2j}(a')^{2n-2j} b^{2j}e^{2it\sum_{l=1}^{2j}\varepsilon(k_{l})}\\+
\sum_{j=0}^{p-1}\binom{2n-1}{2j+1}(a')^{2n-2j-2}b^{2j+2}e^{2it\sum_{l=0}^{2j+1}\varepsilon(k_l)}. 
\end{multline}
It is now clear that when multiplying~\eqref{eq:tr-4} and~\eqref{eq:F-2} the constant term in~\eqref{eq:tr-4} 
can be  treated as in section~\ref{sec:moments}. In fact it gives the same result as~\eqref{eq:final-1} 
after replacing $a\to a'$ and $\ell\to2\ell$. Importantly, there are additional terms that originate from the 
second term in~\eqref{eq:tr-4}. 
First, it is easy to check that the stationary phase approximation gives nonzero $\mu(\zeta_j)$ (cf.~\eqref{eq:measure}) 
only when  terms of~\eqref{eq:tr-4} and~\eqref{eq:F-2} that contain the same quasimomenta $k_j$ are multiplied. 
Notice that the term with $j=0$ in~\eqref{eq:F-2} does not contribute. 
It is also easy to check that within the  stationary phase approximation all the contributions 
give the same $\mu(\{\zeta_j\})\propto \Theta_2(k)$, with  
\begin{equation}
	\label{eq:mu-mi}
	\Theta_2(k):=2|v(k)|t/\ell+\max(2|v(k)|t/\ell,2)-2\max(2|v(k)j|t/\ell,1), \quad v(k)=\varepsilon'(k), 
\end{equation}
where we replaced $k_0=k$. The absolute value $|v(k)|$ originates from the combination of 
the two terms in the sum in~\eqref{eq:tr-4}. 
%
\begin{figure}[t]
\centering
\includegraphics[width=0.75\textwidth]{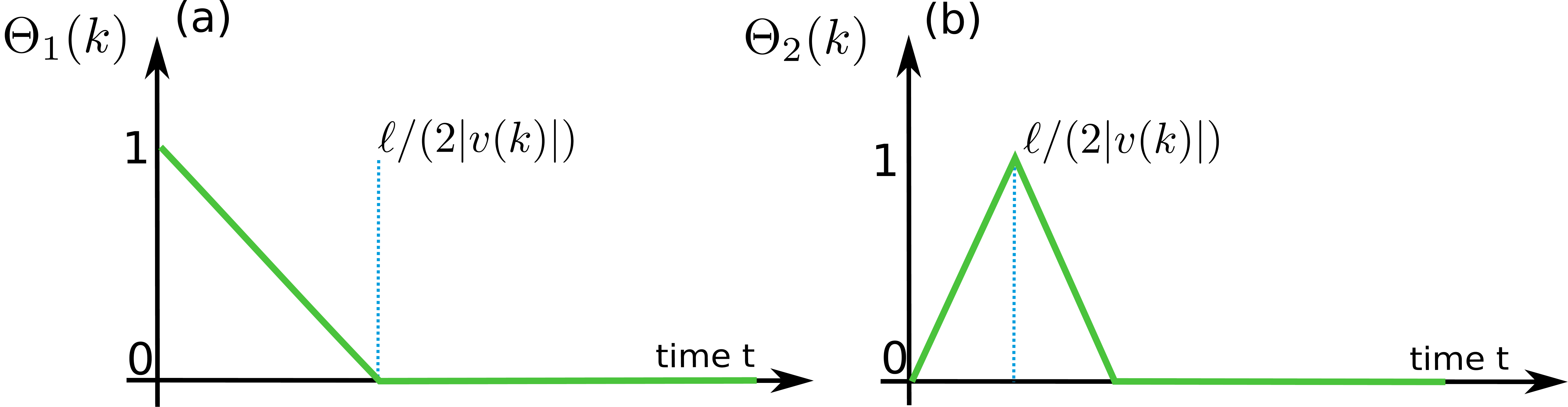}
\caption{ The functions  $\Theta_1(k)=\max(0,1-2|v(k)|t/\ell)$ (a) and 
 $\Theta_2(k)=2|v(k)|t/\ell+\max(2|v(k)|t/\ell,2)-2\max(2|v(k)|t/\ell,1)$ 
 (b) plotted versus time $t$. Here $v(k)=\varepsilon'(k)$ is the fermion group velocity (cf.~\eqref{eq:ham-diag}). 
 The definitions are  for the bipartition with two adjacent intervals of length $\ell$ (see Fig.~\ref{fig0:bip}). 
}
\label{fig:theta}
\end{figure}
%
The function $\Theta_2$ is pictorially defined in Fig.~\ref{fig:theta} (b). 
$\Theta_2(k)$ describes a linear growth up to $t=\ell/(2|v(k)|)$, which is 
followed by a linear decrease up to $t=\ell/(|v(k)|)$. At later times, $\Theta_2(k)$ is 
zero. 
Finally, the sum over $j$ in~\eqref{eq:F-2} gives  
\begin{equation}
	\label{eq:tr-pm-d}
	\mathrm{Tr}[(G^+G^-)^n]=\ell\int_{-\pi}^\pi\frac{dk}{2\pi}\Big(2(a')^{2n}+
	[(a'-b)^{2n}+(a'+b)^{2p}-2(a')^{2n}]\Big(\Theta_1(k)+\frac{1}{2}\Theta_2(k)\Big)\Big). 
\end{equation}
The function $\Theta_1(k)$ is defined as 
\begin{equation}
	\label{eq:theta-1}
\Theta_1(k):=\max(0,1-2|v(k)|t/\ell). 
\end{equation}
$\Theta_1(k)$ is plotted as a function of time in Fig.~\ref{fig:theta} (a). At $t=0$, 
$\Theta_1(k)=1$ and then it decreases linearly up to $t=\ell/(2|v(k)|)$. At later times 
$\Theta_1(k)$ is zero. 
Notice that $\Theta_1(k)+\Theta_2(k)/2=\max(0,\ell-|v(k)|t/\ell)$. 
Thus, it is clear that Eq.~\eqref{eq:tr-pm-d} has the same 
structure as~\eqref{eq:final-0}. Precisely, it is 
twice the result obtained from~\eqref{eq:final-1} after replacing $a\to 2a'$, $b\to2b$, and 
$n\to 2n$. By using the formula 
\begin{equation}
	\label{eq:formula}
	\mathrm{Tr}(\mathds{1}_{2\ell}+G^+G^-)^{-1}=\sum_{p=0}^\infty \mathrm{Tr}(-G^+G^-)^p, 
\end{equation}
we obtain that (cf.~\eqref{eq:gt})  
\begin{multline}
	\label{eq:p-inv}
	\mathrm{Tr}[(\mathds{1}_{2\ell}+G^+G^-)^{-1}]=
	\ell\int_{-\pi}^\pi\frac{dk}{2\pi}\Big\{\frac{2}{1+(a')^{2}}\\-
	\Big[\frac{(a'-b)^{2}}{1+(a'-b)^2}+\frac{(a'+b)^{2}}{1+(a'+b)^2}-\frac{2(a')^{2}}{1+(a')^2}\Big]\Big(\Theta_1(k)+\frac{1}{2}\Theta_2(k)\Big)\Big\}.
\end{multline}
To derive the term in the square brackets in~\eqref{eq:p-inv} one has to remove the term with $p=0$ in~\eqref{eq:formula}. 
This is clear from~\eqref{eq:tr-pm-d} because the term in the square brackets for $p=0$ is zero. 

\subsection{Moments with defects insertions}
\label{sec:m-with-defects} 

Here we provide the formula describing the weak-dissipative hydrodynamic limit of the moments of 
the fermion correlation functions in the presence of defects insertions. 
The result is 
\begin{multline}
	\label{eq:q12-gen-x}
	\mathrm{Tr}\left(\prod_{l=1}^m(G^+G^-)^{q_l}G^{\alpha_l}\right)=
	\ell\int_{-\pi}^\pi\frac{dk}{2\pi}
	\Big\{2 (a')^{2s}+\Big[(a'-b)^{2s}+(a'+b)^{2s}-2(a')^{2s}\Big]\Theta_1\\
	+\frac{1}{2}\Big[(a'-b)^{m+\sum_{l}(2q_{2l-1}-d_{2l-1,2l})}(a'+b)^{\sum_{l}(2q_{2l}+d_{2l-1,2l})}
		+(a'+b\leftrightarrow a'-b)
-2(a')^{2s}\Big]\Theta_2
\Big\}. 
\end{multline}
Eq.~\eqref{eq:q12-gen-x} is obtained from the moments without defects $\mathrm{Tr}(G^+G^-)^n$ 
by inserting the isolated operators $G^{\alpha_l}$. In~\eqref{eq:q12-gen-x} we defined 
$s=m/2+\sum_k q_k$ and $d_{i,j}$ as
\begin{equation}
	\label{eq:d-def}
	d_{i,j}:=\left\{\begin{array}{cc}
		1 & \mathrm{for}\,\,(\alpha_i,\alpha_j)=(+,+)\\
		1 & \mathrm{for}\,\,(\alpha_i,\alpha_j)=(-,-)\\
		0 & \mathrm{for}\,\,(\alpha_i,\alpha_j)=(+,-)\\
		2 & \mathrm{for}\,\,(\alpha_i,\alpha_j)=(-,+)
	\end{array}
	\right.
\end{equation}
The derivation of~\eqref{eq:q12-gen-x} is cumbersome, and we report the main steps in Appendix~\ref{sec:defect}. 
Eq.~\eqref{eq:q12-gen-x} gives access to several other moments constructed from 
the matrices $G^\pm$ and that are needed to obtain the fermionic 
negativity. For instance, by summing over $q_l$ in~\eqref{eq:q12-gen-x} we obtain 
\begin{multline}
\label{eq:t-last-x}
\mathrm{Tr}\left(\prod_{l=1}^m (\mathds{1}_{2\ell}+G^+G^-)^{-1}G^{\alpha_l}\right)=\\
\ell\int_{-\pi}^\pi\frac{dk}{2\pi}
\Big\{2\Big(\frac{a'}{1+(a')^2}\Big)^m+\Big[\Big(\frac{a'-b}{1+(a'-b)^2}\Big)^m
+\Big(\frac{a'+b}{1+(a'+b)^2}\Big)^m
-2\Big(\frac{a'}{1+(a')^2}\Big)^m\Big]\Theta_1\\
+\frac{1}{2}\Big[\frac{(a'+b)^{m-\sum_{l}d_{2l-1,2l}}}{[1+(a'+b)^2]^{m/2}}
\frac{(a'-b)^{\sum_{l}d_{2l-1,2l}}}{[1+(a'-b)^2]^{m/2}}+(a'+b\leftrightarrow a'-b)
-\frac{2(a')^m}{[1+(a')^2]^m}
\Big]\Theta_2
\Big\}. 
\end{multline}
By summing over $\alpha_l$ in~\eqref{eq:t-last-x}, we obtain 
\begin{multline}
	\label{eq:last-x}
	\mathrm{Tr}\Big[\big(G^{\mathrm{T}}\big)^m\Big] =
	\ell\int_{-\pi}^\pi\frac{dk}{2\pi}
	\Big\{\Big(\frac{1}{2}\pm\frac{a'}{1+(a')^2}\Big)^m\\
	+\frac{1}{2}\Big[\Big(\frac{1}{2}\pm\frac{a'-b}{1+(a'-b)^2}\Big)^m
		+\Big(\frac{1}{2}\pm\frac{a'+b}{1+(a'+b)^2}\Big)^m 
		-2\Big(\frac{1}{2}\pm\frac{a'}{1+(a')^2}\Big)^m
\Big]\Theta_1(k)\\
+\frac{1}{2}\Big[\Big(\frac{1}{2}\pm\frac{a'}{[1+(a'+b)^2]^{1/2}[1+(a'-b)^2]^{1/2}}\Big)^m
-\Big(\frac{1}{2}\pm\frac{a'}{1+(a')^2}\Big)^m
\Big]\Theta_2(k)
\Big\}. 
\end{multline}
Here we removed the subscript $A$ in $G^\mathrm{T}_A$ to lighten the notation, although 
the correlation matrices are always restricted to subsystem $A$. 
As it is clear from~\eqref{eq:gt}, Eq.~\eqref{eq:last-x} is 
crucial to compute the logarithmic negativity.

\subsection{Fermionic logarithmic negativity}
\label{sec:neg-res}

We now have all the ingredients to discuss the quasiparticle picture for the fermionic 
logarithmic negativity. Before starting, we observe that~\eqref{eq:s-last} allows one 
to obtain the hydrodynamic behavior of $\mathrm{Tr}[\mathcal{F}(C^\mathrm{T})]$, for 
any $\mathcal{F}(z)$  that admits a Taylor expansion near $z=0$. After expanding 
$\mathcal{F}(z)$, by applying~\eqref{eq:s-last} to all the terms, 
and resummming the series, 
we obtain  that 
\begin{multline}
	\label{eq:last-F}
	\mathrm{Tr}\Big[\mathcal{F}(G^{\mathrm{T}})\Big] =
	\ell\int_{-\pi}^\pi\frac{dk}{2\pi}
	\Big\{\mathcal{F}\Big(\frac{1}{2}\pm\frac{1-a}{1+(1-a)^2}\Big)\\
	+\frac{1}{2}\Big[\mathcal{F}\Big(\frac{1}{2}\pm\frac{1-a-b}{1+(1-a-b)^2}\Big)
		+\mathcal{F}\Big(\frac{1}{2}\pm\frac{1-a+b}{1+(1-a+b)^2}\Big) 
		-2\mathcal{F}\Big(\frac{1}{2}\pm\frac{1-a}{1+(1-a)^2}\Big)
\Big]\Theta_1(k)\\
+\frac{1}{2}\Big[\mathcal{F}\Big(\frac{1}{2}\pm\frac{1-a}{[1+(1-a+b)^2]^{1/2}[1+(1-a-b)^2]^{1/2}}\Big)
-\mathcal{F}\Big(\frac{1}{2}\pm\frac{1-a}{1+(1-a)^2}\Big)
\Big]\Theta_2(k)
\Big\}, 
\end{multline}
where we replaced $a'=1-a$, and where one has to sum over the $\pm$. 
To calculate  the negativity (see first term in~\eqref{eq:gt}) we 
have to use  
\begin{equation}
	\mathcal{F}^{(1/2)}(z):=\ln(z^{1/2}+(1-z)^{1/2}). 
\end{equation}
We also need to calculate $\mathrm{Tr}[\mathcal{F}^{\scriptscriptstyle(2)}(C_A)]$ where 
$C_A$ is the correlation matrix for $A_1\cup A_2$ (cf.~\eqref{eq:c-corr}) of length $2\ell$, with 
\begin{equation}
	\mathcal{F}^{(2)}(z):=\frac{1}{2}\ln(z^2+(1-z)^2). 
\end{equation}
The hydrodynamic prediction for the latter contribution 
is obtained from~\eqref{eq:F-final-1} as 
\begin{multline}
	\mathrm{Tr}[\mathcal{F}^{(2)}(C_A)]=\ell\int_{-\pi}^\pi\frac{dk}{2\pi}[2\mathcal{F}^{(2)}(a/2)\min(1,t/\ell|v(k)|)\\+
	(\mathcal{F}^{(2)}((a+b)/2)+\mathcal{F}^{(2)}((a-b)/2))\max(0,1-t/\ell|v(k)|)]. 
\end{multline}
%
%
\begin{figure}[t]
\centering
\includegraphics[width=0.55\textwidth]{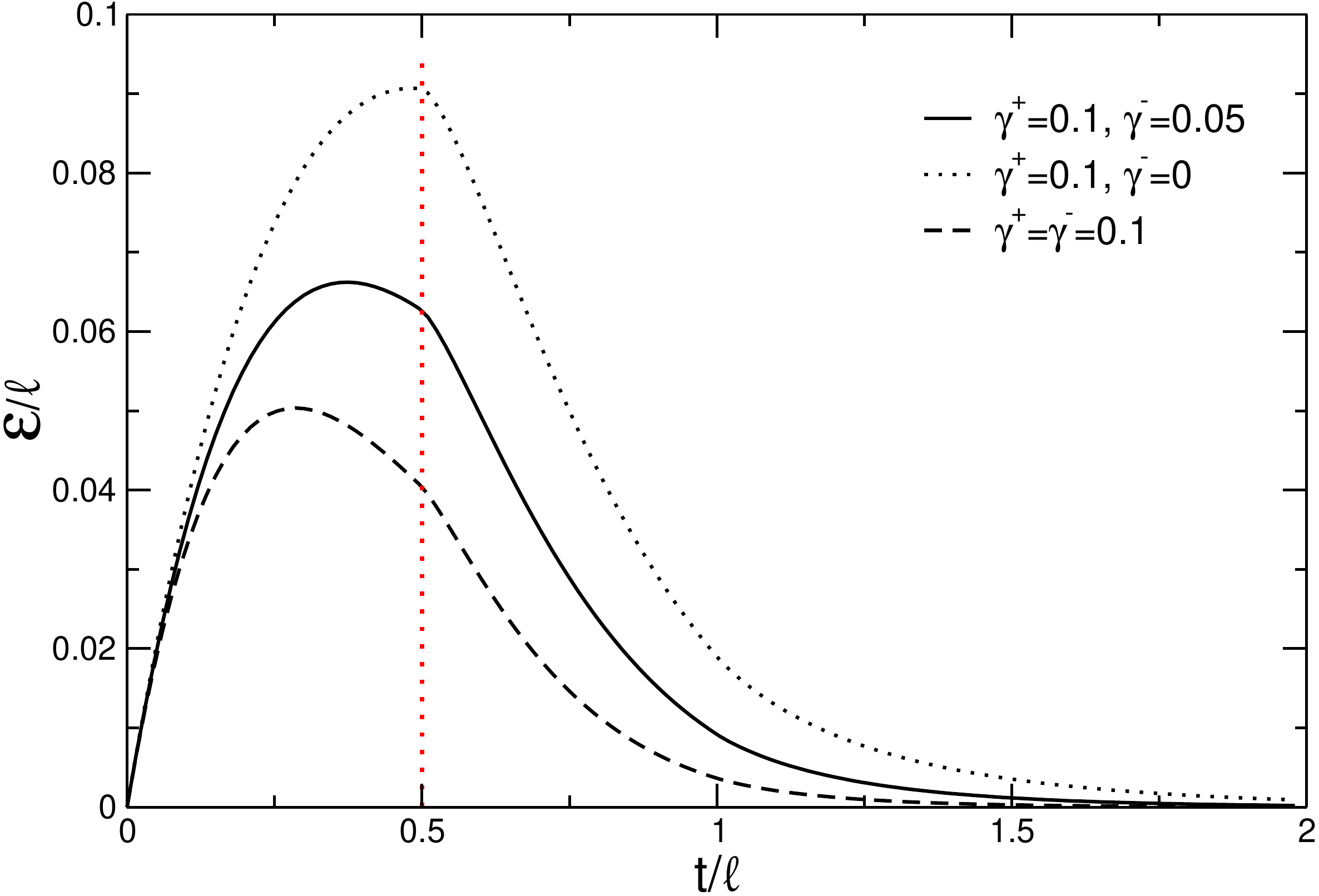}
\caption{Dynamics of the fermionic negativity $\cal{E}$: 
 Theoretical predictions in the weakly-dissipative hydrodynamic limit $t,\ell\to\infty$ 
 $\gamma^\pm\to0$ with $t/\ell$ and $\gamma^\pm\ell$ fixed. The results are 
 for two adjacent intervals of equal length $\ell$. 
 We plot $\cal{E}/\ell$ versus the rescaled time $t/\ell$ for several 
 gain/loss rates $\gamma^\pm$. The negativity is smaller for 
 balanced gain and losses, i.e., for $\gamma^+=\gamma^-$, and it increases 
 upon increasing the inbalance between them. Notice the cusp-like 
 singularity at $t/\ell=1/2$ (marked by the vertical line), 
 which reflects the presence of entangled quasiparticles. 
}
\label{fig:neg_theo}
\end{figure}
%

%
\begin{figure}[t]
\centering
\includegraphics[width=0.55\textwidth]{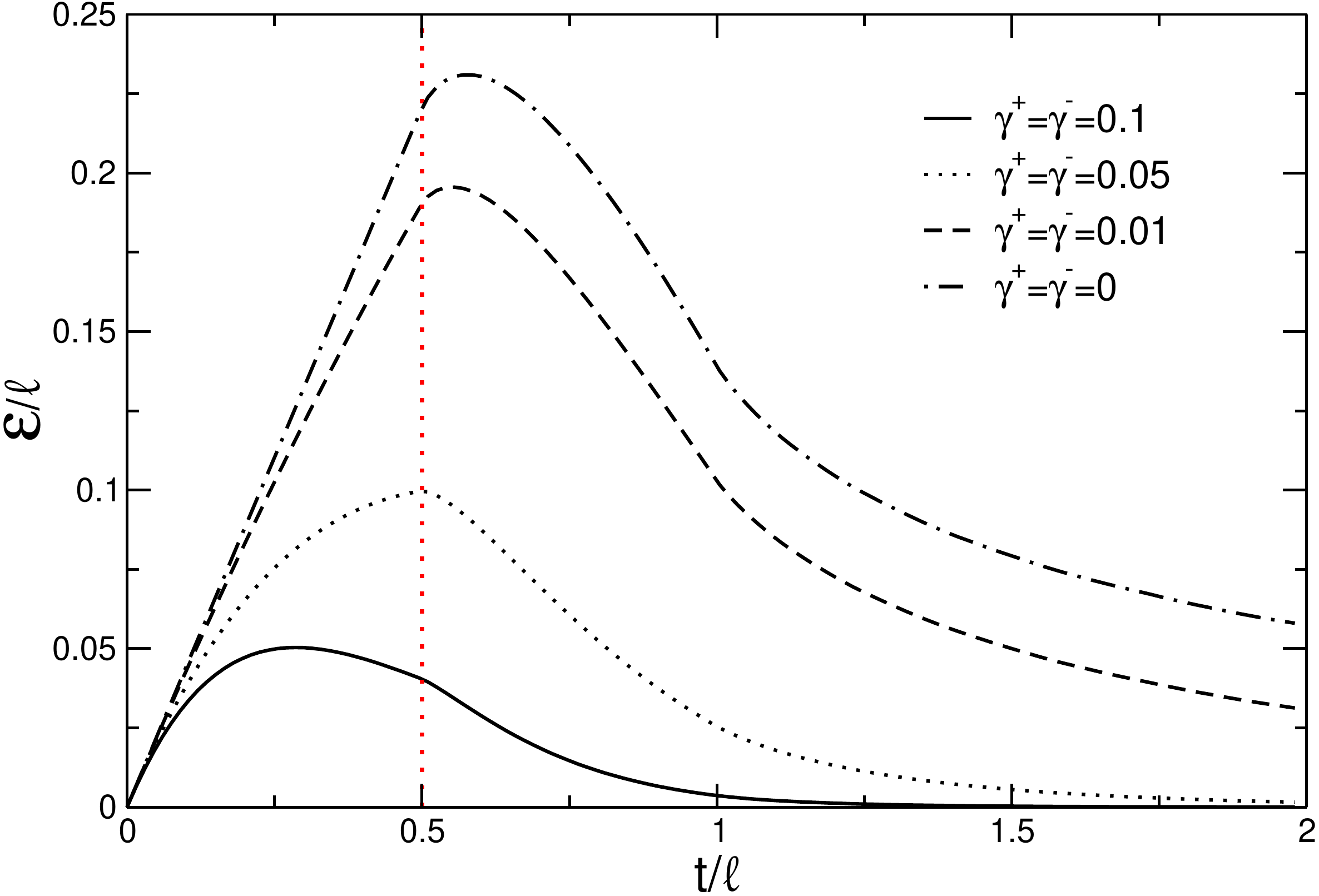}
\caption{Dynamics of the fermionic negativity $\cal{E}$: Same as in 
 Fig.~\ref{fig:neg_theo} for balanced gain and losses, i.e., $\gamma^+=\gamma^-$. 
 We show the theoretical prediction for the negativity for $\gamma^+=0.1,0.05,0.01$. 
 The dashed-dotted line is the prediction in the absence of dissipation, i.e., for $\gamma^+=\gamma^-=0$. 
 In the absence of dissipation $\cal{E}$ exhibits a linear growth up to $t/\ell=1/2$, followed by 
 a ``slow'', i.e., power-law decay. Upon switching on the dissipation the initial linear 
 growth is damped. Approximate linear growth can be observed only for $t\ll 1/\gamma^\pm$. 
 Moreover, for finite $\gamma^\pm$, ${\cal E}$ vanishes exponentially at $t/\ell\to\infty$. 
}
\label{fig:neg_theo_supp}
\end{figure}
%

Let us consider the unitary limit $\gamma^\pm\to0$. Thus, one 
obtains that $a\to1$ and $b\to1$. From~\eqref{eq:last-F}, one has that 
\begin{equation}
	\label{eq:neg-1}
	\mathrm{Tr}(\mathcal{F}^{(1/2)}(G_A^\mathrm{T}))=2\ell\int_{-\pi}^\pi\frac{dk}{2\pi}\mathcal{F}(1/2)[1-\Theta_1(k)],  
\end{equation}
where we used that $\mathcal{F}^{\scriptscriptstyle(1/2)}(0)=\mathcal{F}^{\scriptscriptstyle(1/2)}(1)=0$. 
Importantly, as we also observed before,  the term 
with $\Theta_2$ vanishes in the non-dissipative limit. On the other hand,  we have 
\begin{equation}
	\label{eq:neg-2}
	\mathrm{Tr}(\mathcal{F}^{(2)}(C_A))=2\ell\int_{-\pi}^\pi\frac{dk}{2\pi}\mathcal{F}^{(2)}(1/2)\min(1,|v(k)|t/\ell). 
\end{equation}
Putting together~\eqref{eq:neg-1}~\eqref{eq:neg-2} and~\eqref{eq:gt}, we obtain that 
\begin{equation}
	\label{eq:neg-unit}
	{\cal E}=\frac{\ell}{2}\int_{-\pi}^\pi\frac{dk}{2\pi}\ln(2)\Theta_2(k). 
\end{equation}
This is in agreement with the conjecture that 	${\cal E}=I^{(1/2)}_{A_1:A_2}/2$ 
that was put forward in~\cite{alba2019quantum}, 
where $I^{\scriptscriptstyle(1/2)}_{A_1:A_2}$ is the R\'enyi mutual information with 
R\'enyi index $1/2$ between the two intervals. This conjecture was also verified recently 
for the quench from the N\'eel state and the Majumdhar-Ghosh state in the $XX$ chain~\cite{parez2022dynamics}. 
As we are going to discuss, the relation between negativity and R\'enyi mutual information 
is violated in the presence of dissipation. 

Before discussing the final result for $\mathcal{E}$ we observe that 
\begin{equation}
	\label{eq:supp}
	\mathcal{F}^{(1/2)}\left(\frac{1}{2}+\frac{1-a}{1+(1-a)^2}\right)
	+\mathcal{F}^{(1/2)}\left(\frac{1}{2}-\frac{1-a}{1+(1-a)^2}\right)
	+2\mathcal{F}^{(2)}\left(\frac{a}{2}\right)=0. 
\end{equation}
The left hand side of Eq.~\eqref{eq:supp} would be the contribution to ${\cal E}$ for $t>\ell/|v(k)|$. 
Eq.~\eqref{eq:supp} is consistent with the fact  that the negativity is determined by 
the propagation of entangled pairs of quasiparticles. Indeed, for $t>\ell/|v(k)|$ there 
are no entangled pairs shared between the two subsystems. 
Furthermore, within the quasiparticle picture the negativity should be proportional 
to $\Theta_2$, which is the number of entangled pairs that are shared between 
$A_1$ and $A_2$, as in the unitary case (cf.~\eqref{eq:neg-unit}). Indeed, a straightforward 
calculation shows that the negativity is given by 
\begin{equation}
	\label{eq:neg-fin}
	{\cal E}=\frac{\ell}{2}\int_{-\pi}^\pi\frac{dk}{2\pi}e(k)\Theta_2(k), 
\end{equation}
where
\begin{equation}
	\label{eq:neg-den}
e:=\ln\left(\frac{1}{2}\left(1-(1-a)^2+b^2+\sqrt{[1+(1-a)^2]^2+2[1-(1-a)^2]b^2+b^4}\right)\right) 
\end{equation}
is the density of negativity. 
As anticipated, the function $\Theta_2(k)$ appears in~\eqref{eq:neg-fin}. 
	The structure of~\eqref{eq:neg-fin} is quite similar to the result for the 
	logarithmic negativity after a global quantum quench in Conformal Field Theory~\cite{coser2014entanglement}. 
	Indeed, the same function $\Theta_2(k)$ appears. This gives the typical ``rise and fall'' dynamics of 
	the negativity. Specifically, for two adjacent intervals of equal length $\ell$, 
	${\cal E}$ grows linearly up to time $t/(2v_\mathrm{max})$, with $v_\mathrm{max}$ the maximum 
	velocity in the system. At asymptitocally long times the negativity vanishes. In contrast with the 
	behavior in Conformal Field Theory where the negativity decreases linearly up to $t=\ell/v_\mathrm{max}$, 
	where it vanishes, Eq.~\eqref{eq:neg-fin} predicts a ``slow'' vanishing behavior. This is due to the fact that 
	for lattice models the quasiparticles possess a nonlinear dispersion. The nonzero negativity at 
	times $t>\ell/v_\mathrm{max}$ is due to slow quasiparticles. 

	Let us now discuss the negativity content of the quasiparticle pairs $e(k)$ in~\eqref{eq:neg-fin}. First, we 
	should observe that $e(k)$ depends on time because $a$ and $b$ are time-dependent (cf.\eqref{eq:C-map}~\eqref{eq:t-prime}). 
	This is different from the CFT setup~\cite{coser2014entanglement} and from the case of free-boson and free-fermion 
	systems~\cite{alba2019quantum}. As a consequence of the factor $e^{-(\gamma^++\gamma^-)t}$ in the 
	definition of $b$ (cf.~\eqref{eq:C-map}) in the presence of dissipation the linear growth at $t\le \ell/(2v_\mathrm{max})$ 
	is damped. Indeed, approximate linear behavior is visible only for $t\ll 1/(\gamma^++\gamma^-)$. Moreover, 
	at long times $t\gg \ell$ the negativity decays to zero exponentially, in constrast with unitary dynamics, for which 
	the decay is power law. 
	Furthermore, in integrable free-fermions and free-boson systems, the negativity content of the 
	quasiparticle pairs is the R\'enyi mutual information with R\'enyi index $1/2$. As it is clear from~\eqref{eq:neg-fin}, 
	this is not the case in the presence of dissipation.

Clearly, from~\eqref{eq:neg-fin} one recovers that 
in the non-dissipative case $\gamma^\pm=0$, $e=\ln(2)$. 
Finally, it is interesting to focus on the balanced gain/loss dissipation. 
The condition $\gamma^+=\gamma^-$ implies that $a=1$, whereas $b=e^{-2\gamma^-t}$. 
This means that the term proportional to $\Theta_2(k)$ in~\eqref{eq:last} vanishes. 
Now, one has that $e(k)=\ln(1+b^2)$. Interestingly, this implies that 
Eq.~\eqref{eq:neg-den} is 
\begin{equation}
	e(k)=s_k^{(2),\mathrm{YY}}(t)-s_k^{(2),\mathrm{mix}}(t), \quad\mathrm{for}\,\, \gamma^+=\gamma^-,
\end{equation}
where $s_k^{\scriptscriptstyle(2),\mathrm{YY}}$ and $s^{\scriptscriptstyle(2),\mathrm{mix}}$ 
are the same as in~\eqref{eq:ent-quasi}. This means that 
\begin{equation}
	\label{eq:I2}
	{\cal E}=\frac{1}{2}I_{A_1:A_2}^{(2)},\quad\mathrm{for}\,\,\gamma^+=\gamma^- 
\end{equation}
which makes apparent that in dissipative settings ${\cal E}\ne I^{\scriptscriptstyle(1/2)}/2$. 
We also verified that Eq.~\eqref{eq:I2} does not remain valid for generic 
gain and loss processes. 

One should also stress that the structure of~\eqref{eq:neg-fin} is quite 
revealing. Indeed, it is clear that Eq.~\eqref{eq:neg-fin} can be interpreted as the negativity 
of a few qubit systems. It should be possible to derive an effective few-qubits mixed density 
matrix describing the state of $A_1\cup A_2$ that give the negativity in~\eqref{eq:neg-fin}. 
Importantly, the effective density matrix is expected to be mixed because of the dissipative 
dynamics. The fact that the dynamics is dissipative is at the heart of the failure of the result 
of Ref.~\cite{alba2019quantum}, which relies on the local dynamics being unitary.  
This idea allows to understand the dynamics of the negativity free-fermions in the presence of 
localized losses~\cite{caceffo2022entanglement}. Still, to derive the effective density matrix 
for the two intervals one would need at least a different quench in order to be able to guess 
how the elements of the matrix depend on the parameters of the system.

Our theoretical predictions for balanced gain and loss dissipation are reported in 
	Fig.~\ref{fig:neg_theo_supp}. In the figure we show results for vanishing dissipation 
	rates $\gamma^+=\gamma^-=0.1,0.05,0.01,0$. Clearly, as the dissipation is switched off, 
	the unitary result is recovered. In particular, one recovers the linear increase up to 
	$t={\mathcal O}(\ell)$. 

Finally, within the quasiparticle picture it 
is straightforward to generalize~\eqref{eq:neg-fin} to the case of two intervals at 
a distance $d$ (see Fig.~\ref{fig0:bip}). Indeed, as for the unitary case~\cite{alba2019quantum}, 
it is natural to expect that the negativity content $e(k)$ of the entangled pairs remains the same 
as in~\eqref{eq:neg-fin}, whereas only the function $\Theta_2(k)$ has to be modified. 
This is obtained by replacing $\Theta_2(k)$ in~\eqref{eq:neg-fin} with $\widetilde\Theta_2(k)$ 
defined as 
\begin{equation}
	\label{eq:theta-dis}
\widetilde\Theta_2:=\max(2|v(k)|t/\ell,2+d/\ell)+  
\max(2|v(k)|t/\ell,d/\ell)-
2\max(2|v(k)|t/\ell,1+d/\ell). 
\end{equation}
Clearly, $\widetilde\Theta_2(k)$ appears in the quasiparticle picture for the mutual 
information between two intervals (see for instance~\cite{murciano2021quench}). This 
happens because both the mutual information and the negativity are proportional to the 
number of pairs shared between $A_1$ and $A_2$. 
$\widetilde\Theta_2(k)$ is zero for $t<d/(2|v(k)|)$. This reflects that at very short  
times there are entangled pairs that are shared between $A_1\cup A_2$ and the rest, but there 
are no entangled pairs shared between $A_1$ and $A_2$ only. $\widetilde\Theta_2(k)$ grows linearly 
for $d/(2|v(k)|)\le t<(d+\ell)/(2|v(k)|)$. At later times $\widetilde\Theta_2(k)$ 
decreases linearly. At any $t>(d+2\ell)/(2|v(k)|)$ it is identically zero. 

We report analytical predictions for $\cal{E}$ for several dissipation rates $\gamma^\pm$ 
in Fig.~\ref{fig:neg_theo} considering the case of adjacent intervals. 
For all the values of $\gamma^\pm$ the negativity exhibits 
the typical ``rise and fall'' behavior, with a growth at short times, followed by a 
vanishing behavior in the long time 
limit. The maximum at intermediate times is lower  for the case with 
balanced gain/loss, i.e., for $\gamma^+=\gamma^-$, and it progressively grows as the 
imbalance is increased. The vertical dashed line marks the point $t/\ell=1/2$. All the 
curves exhibit a cusp-like singularity at this point, which reflects the presence of 
the entangled quasiparticles. Indeed, a similar cusp is present in the absence of 
dissipation. 

Finally, let us stress again that  Eq.~\eqref{eq:neg-fin} and Eq.~\eqref{eq:neg-den} 
hold in the usual hydrodynamic limit with $\ell,d,t\to\infty$ 
with the ratios $t/\ell$ and $t/d$ fixed. Still, since ${\cal E}$ at fixed $\gamma^\pm$ vanishes 
exponentially as $e^{-(\gamma^++\gamma^-)t}$ for $t\to\infty$, it is convenient to take the weakly-dissipative hydrodynamic limit 
by sending $\gamma^\pm\to0$ with fixed $\gamma^\pm\ell$.  

\section{Numerical benchmarks}
\label{sec:numerics}

Having derived the quasiparticle picture for the logarithmic negativity in the (weakly-dissipative) 
hydrodynamic limit, we now discuss some numerical checks. We first focus on several moments 
of the matrices $G^+G^-$ in section~\ref{sec:mom-num}. Finally, in section~\ref{sec:num-neg} 
we discuss numerical results for the negativity.

\subsection{Moments of $G^+G^-$}
\label{sec:mom-num}

\begin{figure}[t]
\centering
\includegraphics[width=0.55\textwidth]{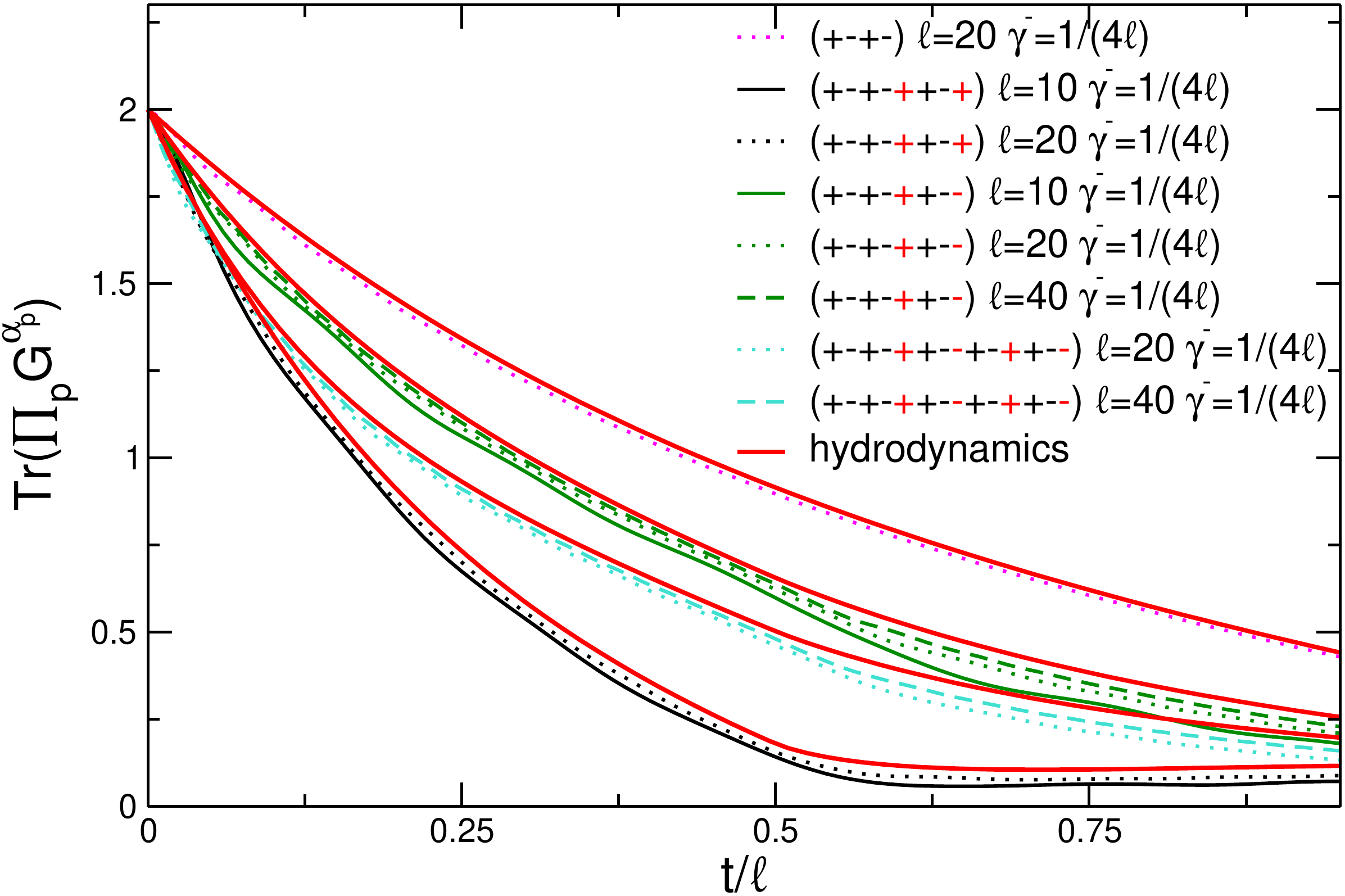}
\caption{ Dynamics of the moments $\mathrm{Tr}(\prod_p G^{\alpha_p})$ in the 
 tight-binding chain with gain and loss dissipation. The results are for the quench from 
 the fermionic N\'eel state and for two adjacent intervals of equal length $\ell$ (see Fig.~\ref{fig0:bip}). 
 The string of $(\alpha_1,\alpha_2,\dots)$ that identifies the correlator is 
 reported. 
 Here we consider strings of operators $G^\pm$ with and without ``defects'', 
 i.e., places where the same operator appears on consecutive sites (red symbols). 
 We fix $\gamma^+=0$ and $\gamma^-=1/(4\ell)$. 
 The continuous red lines are the analytical results in the weakly-dissipative hydrodynamic 
 limit $\ell,t\to\infty$ with $t/\ell$ and $\gamma^-\ell$ fixed. The theoretical results 
 are obtained by using Eq.~\eqref{eq:tr-pm-d} and Eq.~\eqref{eq:q12-gen-x} 
}
\label{fig4:check_def}
\end{figure}

Let us discuss the moments 
\begin{equation}
	\label{eq:corr-num}
	M_n:=\mathrm{Tr}\Big[\prod_{p=1}^nG^{\alpha_p}\Big],\quad\mathrm{with}\,\,\alpha_p=\pm. 
\end{equation}
Here we focus on two adjacent intervals of length $\ell$. 
The correlator~\eqref{eq:corr-num} is identified by the 
string $\{\alpha_1,\alpha_2,\dots,\alpha_n\}$. Our numerical results for 
$M_n$ are shown in Fig.~\ref{fig4:check_def}. 
%
%
We only consider the situation with loss dissipation with rate $\gamma^-$. 
In Fig.~\ref{fig4:check_def}  we consider both moments with 
defects insertions (see section~\ref{sec:defect}), as well as without them. 
The operator insertions that create defects are denoted with red $\pm$ symbols. For each 
$M_n$ we consider several values of increasing $\ell$. To reach the weakly-dissipative 
hydrodynamic limit we fix $\gamma^-=1/(4\ell)$. The analytic results in the 
scaling limit are reported as continuous red lines, and are obtained by 
using~\eqref{eq:q12-gen-1}. For all the cases that we consider, at $t=0$ 
we have $M_n=2$ independently of $n$. Then, $M_n$ decrease, vanishing in the 
limit $t\to\infty$. It is important to stress that this is due to the fact that 
we have only loss dissipation. In the generic case with  both gain and loss dissipation 
the behavior is different. Precisely, the moments start at $M_n=2$, they exhibit a 
minimum at intermediate times, and saturate to a nonzero value at $t\to\infty$. 
As it is clear from Fig.~\ref{fig4:check_def},  
as we approach the weakly-dissipative hydrodynamic limit, deviations between 
the exact numerical data and the analytic predictions become 
progressively smaller. 
%
\begin{figure}[t]
\centering
\includegraphics[width=0.55\textwidth]{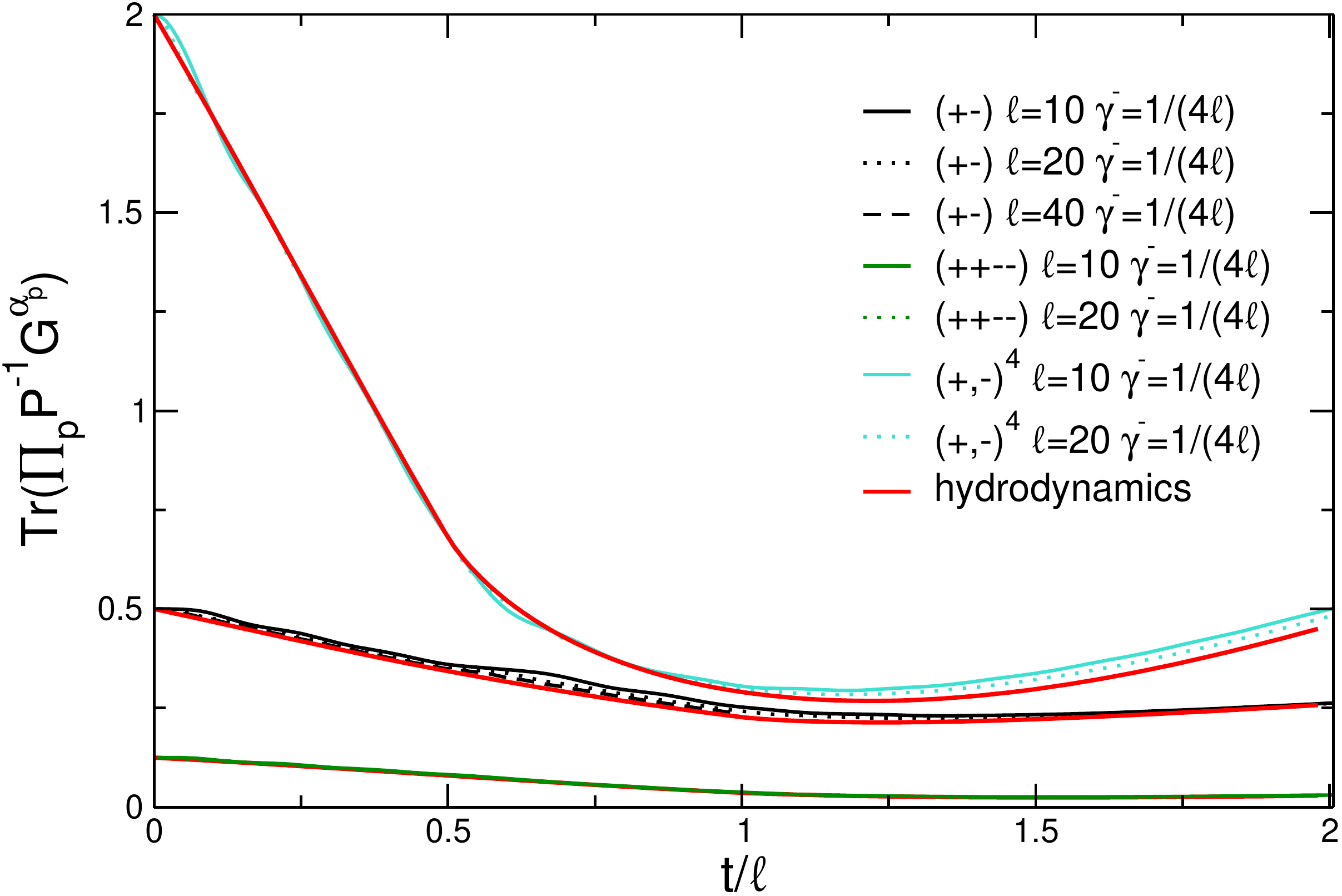}
\caption{Dynamics of the moments $\mathrm{Tr}(\prod_p P^{-1}G^{\alpha_p})$ with $\alpha_p=\pm$ 
 and $P=\mathds{1}_{2\ell}+G^+G^-$. Results are for the quench from the fermionic N\'eel state and 
 for two adjacent intervals of equal length $\ell$. 
 The gain and loss rates $\gamma^\pm$ are fixed as  $\gamma^+=0$ and $\gamma^-=1/(4\ell)$. 
 In the legend we report as $(\alpha_1,\alpha_2,\dots)$ the configuration of the $\alpha_p$ 
 that identify the different moments. Here $(+,-)^4$ denotes the sum over all the possible 
 strings $(\alpha_{1},\alpha_2,\alpha_3,\alpha_4)$. The continuous red lines are the 
 results in the weakly-dissipative hydrodynamic limit $\ell,t\to\infty$ with $t/\ell$ and 
 $\gamma^\pm\ell$ fixed. The analytical results are given by Eq.~\eqref{eq:t-last-x}. 
}
\label{fig5:checkP}
\end{figure}
%
In Fig.~\ref{fig5:checkP} we consider the moments $M'_n$ defined as 
\begin{equation}
	\label{eq:mom-1}
	M'_n:=\mathrm{Tr}\Big[\prod_{p=1}^n(\mathds{1}_{2\ell}+G^+G^-)^{-1}G^{\alpha_p}\Big]. 
\end{equation}
Similar to Fig.~\ref{fig4:check_def}, 
we focus on $\gamma^+=0$. The red continuous lines are the results in the 
weakly-dissipative hydrodynamic limit. These are obtained by using~\eqref{eq:t-last}. 
Already for moderately small $\gamma^-$ and large $t,\ell$ the data are in very good 
agreement with the analytical results. 
%
\begin{figure}[t]
\centering
\includegraphics[width=0.55\textwidth]{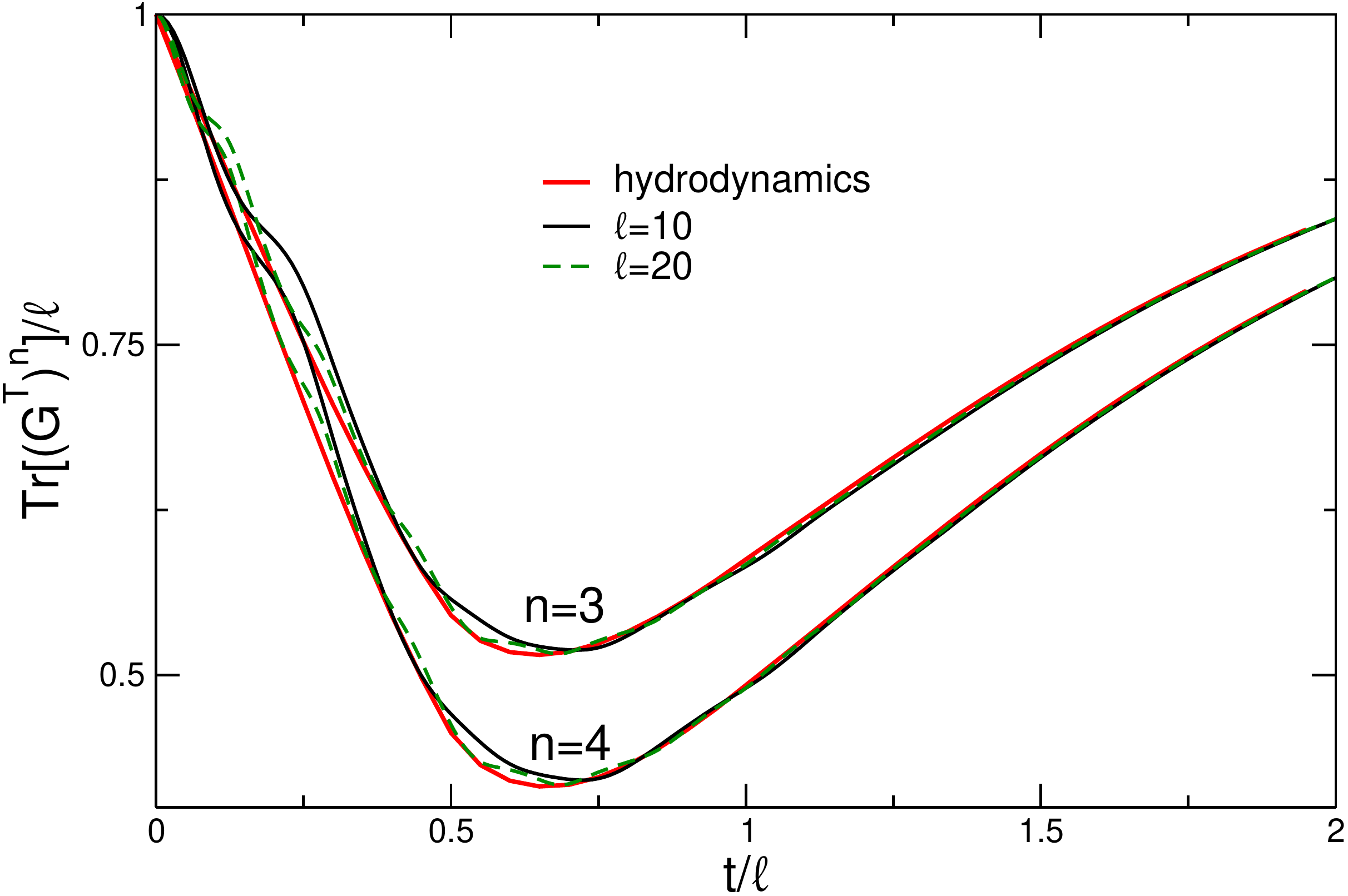}
\caption{Dynamics of the moments $\mathrm{Tr}[(G^T)^n]$ in the tight-binding chain 
 with gain and loss dissipation. The results are for two adjacent intervals of equal 
 length $\ell$ (see Fig.~\ref{fig0:bip}). Dissipation rates $\gamma^\pm$ are chosen as 
 $\gamma^+=1/(2\ell)$ and $\gamma^-=0$. We show the rescaled moments $\mathrm{Tr}(G^T)^n/\ell$ 
 versus $t/\ell$. The results are for $n=2,3$ and $\ell=10,20$. The continuous red line is 
 the the weakly-dissipative hydrodynamic limit $t,\ell\to\infty$, with $t/\ell$ and 
 $\gamma^\pm\ell$ fixed.  The analytical results are given by Eq.~\eqref{eq:last-x}. 
}
\label{fig6:GT}
\end{figure}
%
As a further check, in Fig.~\ref{fig6:GT} we discuss the moments of $G^\mathrm{T}$ (cf.~\eqref{eq:gt}). 
We report $\mathrm{Tr}[(G^{\mathrm{T}})^n]$ for $n=3,4$. We now consider gain dissipation only with $\gamma^+=1/(2\ell)$ 
and $\gamma^-=0$. As it is clear from Fig.~\ref{fig6:GT}, already for $\ell=10,20$ the data 
are in excellent agreement with the analytical results in the weakly-dissipative hydrodynamic 
limit (continuous red lines) obtained from~\eqref{eq:last}. 

\subsection{Logarithmic negativity}
\label{sec:num-neg}

\begin{figure}[t]
\centering
\includegraphics[width=0.55\textwidth]{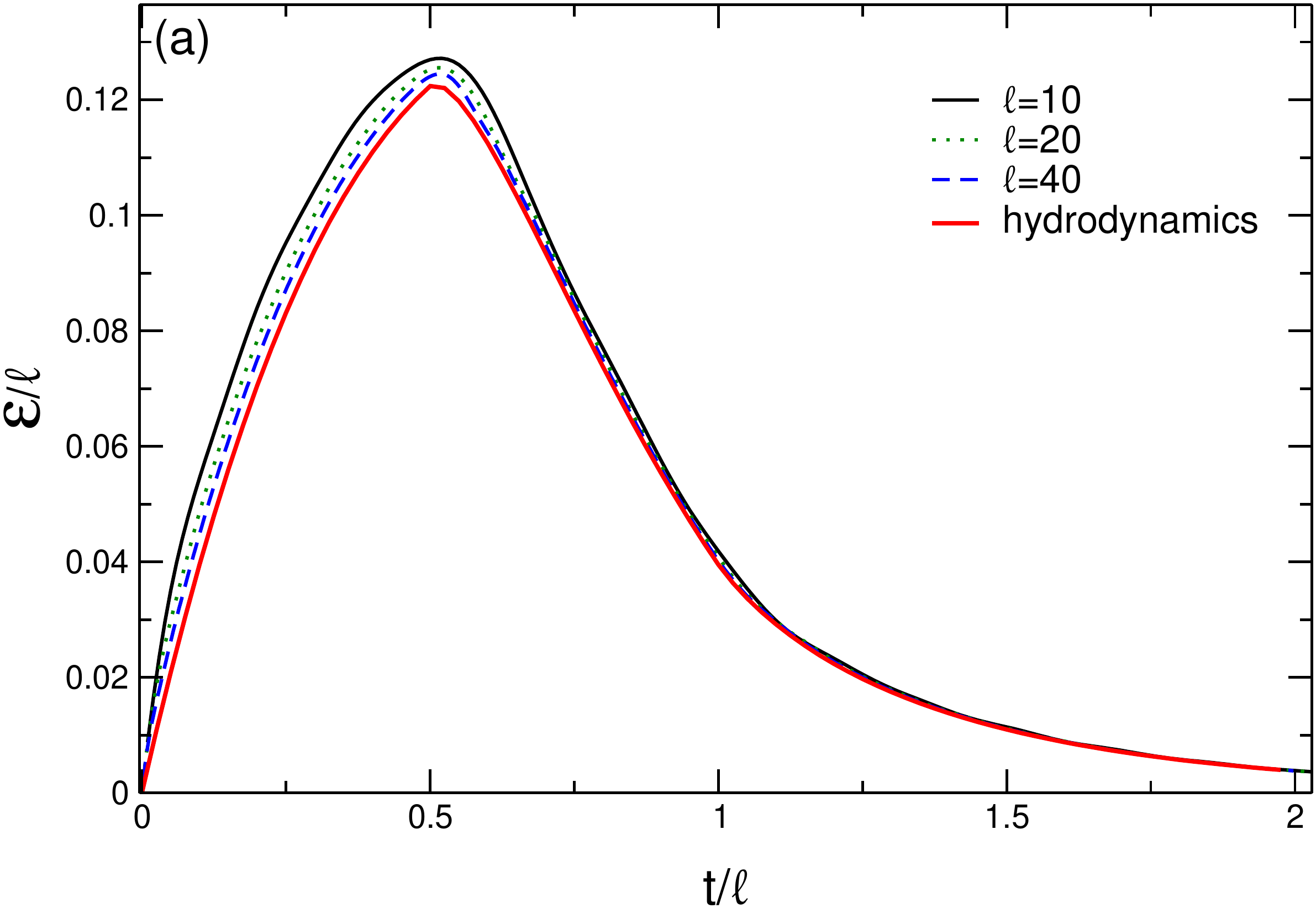}
\includegraphics[width=0.55\textwidth]{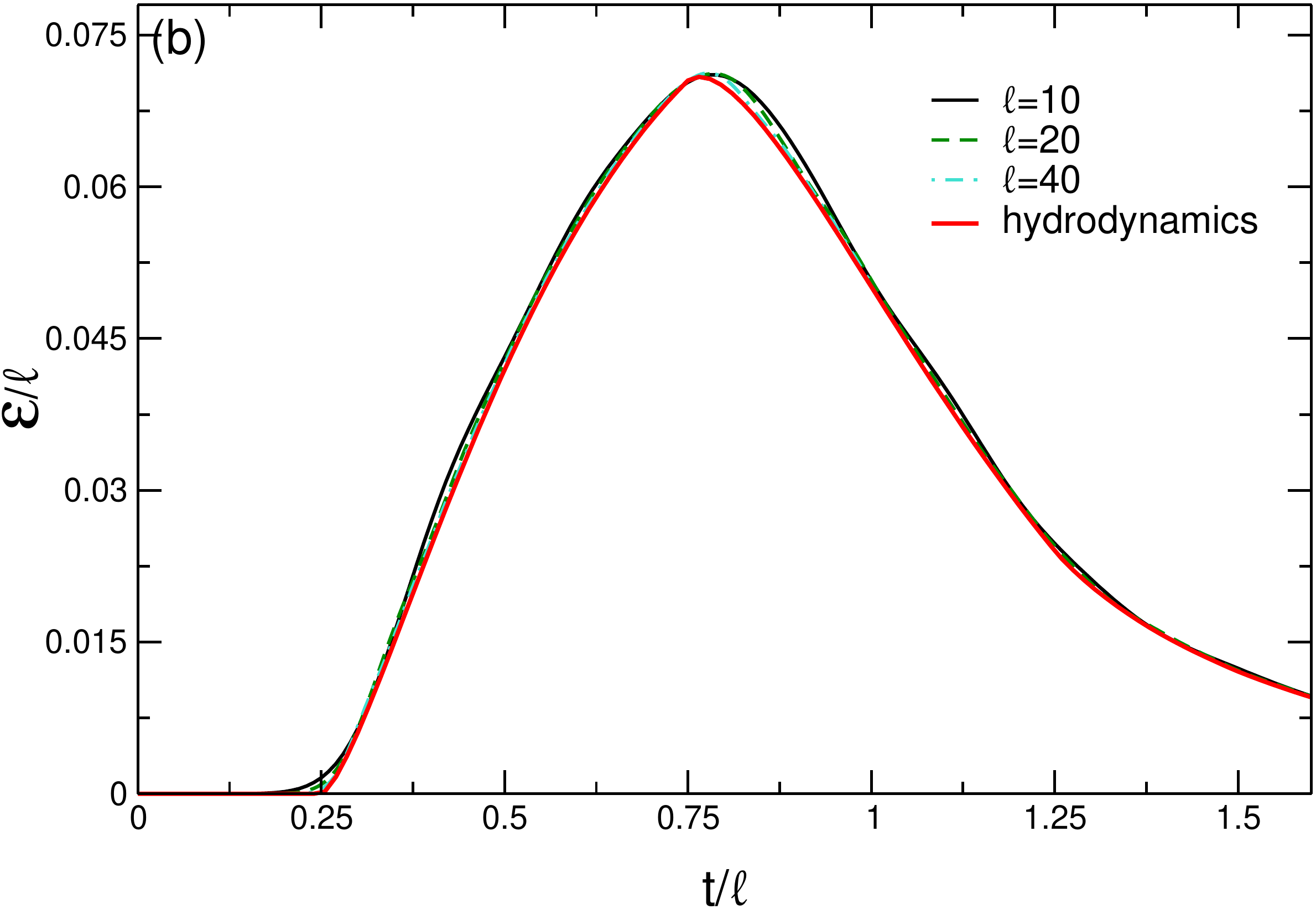}
\caption{Dynamics of the fermionic negativity $\cal{E}$ after the quench 
 from the fermionic N\'eel state in the tight-binding chain with gain/loss dissipation. 
 Results are for the negativity between two intervals of equal length $\ell$. 
 The data are for $\gamma^+=1/(2\ell)$ and $\gamma^-=\gamma^+/2$. The figure 
 shows the scaling plot of $\cal{E}/\ell$ versus $t/\ell$. The continuous red line 
 is the result in the weakly-dissipative hydrodynamic limit. In (a) we show results 
 for two adjacent intervals, whereas in (b) we discuss the case of two disjoint intervals 
 at $d=\ell/2$. The analytical results are given by~\eqref{eq:neg-fin}, where 
 for disjoint intervals we replaced $\Theta_2\to\widetilde\Theta_2$ (cf.~\eqref{eq:theta-dis}). 
}
\label{fig7:neg}
\end{figure}

Let us finally discuss the dynamics of the fermionic negativity. We show numerical 
data for the rescaled fermionic negativity ${\cal E}/\ell$ plotted versus $t/\ell$ 
in Fig.~\ref{fig7:neg}. 
We now consider both gain and loss dissipation with rates $\gamma^+=1/(2\ell)$ and $\gamma^-=\gamma^+/2$. 
In Fig.~\ref{fig7:neg} (a)  we consider the situation with two adjacent intervals 
of equal length $\ell$, i.e., at distance $d=0$ (see Fig.~\ref{fig0:bip}). In 
Fig.~\ref{fig7:neg} (b) we focus on two disjoint intervals. 
Since we are interested in the hydrodynamic limit, we consider $d=\ell/2$. 
%
%
In the figure we show numerical data for $\ell=10,20,40$. The data exhibit the 
typical ``rise and fall'' dynamics. For the two disjoint intervals (Fig.~\ref{fig7:neg}), 
${\cal E}=0$ for $t\le d/(2v_\mathrm{max})$ with $v_\mathrm{max}=1$. This is expected 
because for $t\le d/(2v_\mathrm{max})$ there are no pairs of entangled quasiparticles that 
are shared between $A_1$ and $A_2$. Indeed, the first entangled pair contributing to the 
entanglement between the two intervals is created at a distance $d/2$ from them. The time 
$d/(2v_\mathrm{max})$ is the time at which the two quasiparticles forming the pair 
and traveling with $|v_\mathrm{max}|=1$ reach $A_1$ and $A_2$, respectively. 
The quasiparticle prediction for the logarithmic negativity (cf.~\eqref{eq:neg-2}) is 
reported in Fig.~\ref{fig7:neg} as continuous red line. The agreement between~\eqref{eq:neg-2} 
and the numerical data is remarkable for both adjacent and disjoint intervals. 

\section{Conclusions}
\label{sec:concl}

We derived an exact formula for the dynamics of the fermionic logarithmic negativity 
after the quench from the fermionic N\'eel state in the tight-binding chain with both 
gain and loss dissipation. Our main result is formula~\eqref{eq:neg-fin}. 
As a byproduct we provided analytical results for several 
fermionic correlators.  Formula~\eqref{eq:neg-fin} 
shows that the negativity admits a quasiparticle picture interpretation. 
Similar to the mutual information, the negativity is proportional to the number of entangled 
pairs that are shared between two intervals. This is reflected in its typical ``rise and fall''
dynamics. Still, the negativity content of the quasiparticles originates from an intricate 
interplay between unitary and dissipative contributions. In particular, the negativity  
is not easily related to standard thermodynamic quantities. 
This is in constrast with what happens for the mutual information, which can 
be related to the thermodynamic entropy of the system~\cite{carollo2022dissipative,alba2022hydrodynamics}. 
Moreover, our result shows explicitly 
that in the presence of dissipation the logarithmic negativity is not half of the R\'enyi mutual 
information with R\'enyi index $1/2$, in contrast with the unitary case~\cite{alba2019quantum}. 

Let us now mention some interesting future directions. First, it would be important to 
extend the quasiparticle picture for the negativity to other quenches and other free-fermion 
systems. Indeed, it is likely that a formula for generic quenches and 
quadratic dissipation can be obtained. A good starting point would be to consider the 
quench from the Majumdar-Ghosh state in the tight-binding chain~\cite{parez2021exact}. 
Another interesting direction would be to consider out-of-equilibrium dynamics in 
bosonic systems~\cite{carollo2022dissipative}. Moreover, it would be interesting to 
study the negativity in the presence of localized dissipations. Indeed, it has been 
shown in Ref.~\cite{alba2021unbounded} that the dynamics of the von Neumann and the 
R\'enyi entropies in the presence of localized fermion losses are determined by 
the effective transmission and reflection coefficients of the lossy site. 
It would be interesting to understand how to generalize this result to the negativity. 
Finally, an important open problem is to understand the behavior of the logarithmic 
negativity in dissipative interacting integrable systems. 

\section*{Acknowledgements}
F.C.~acknowledges support from the “Wissenschaftler-R\"uckkehrprogramm GSO/CZS” of the Carl-Zeiss-Stiftung and the German Scholars Organization e.V., as well as through the Deutsche Forschungsgemeinsschaft (DFG, German Research Foundation) under Project No. 435696605, as well as through the Research Unit FOR 5413/1, Grant No. 465199066. F.C.~is indebted to the Baden-W\"urttemberg Stiftung for the financial support by the Eliteprogramme for Postdocs.


\appendix

\section{Moments of $G^+G^-$ with insertions of defects}
\label{sec:defect}

In order to calculate the moments of $G^{\mathrm T}$ (cf.~\eqref{eq:gt}) one has to 
deal with terms of the form 
\begin{equation}
	\label{eq:D}
	\mathrm{Tr}\left(\prod_{l=1}^m(G^+G^-)^{q_l}
	G^{\alpha_l}\right),\quad\mathrm{with}\,\,\alpha_l=\pm. 
\end{equation}
Here $q_l$ is a positive integer. The term $(G^+G^-)^{q_l}$ is obtained by using~\eqref{eq:formula} 
to expand $(\mathds{1}_{2\ell}+G^+G^-)^{-1}$.  The term under the trace in~\eqref{eq:D} is obtained by breaking 
the alternating pattern in $(G^+G^-)^{\sum_{l=1}^{m} q_l}$ via the insertion of 
$m$ ``misplaced'' matrices (defects) $G^{\alpha_l}$  
at positions $2q_l+l$. Now, following the approach in Section~\ref{sec:pm}, one obtains 
an expression similar to~\eqref{eq:F-1}. 
In particular, the presence of the defects in~\eqref{eq:D} 
does not affect the second trace 
in~\eqref{eq:F-1}, which depends only on the details of the quench and of the Hamiltonian. 
The term inside the first trace in~\eqref{eq:F-1} has to be modified, although in a simple manner. 
Specifically, some of the terms $e^{ik_j\ell}+e^{ik_{j+1}\ell}$ in 
the product in~\eqref{eq:idpm} get a relative minus sign. 

Before considering the generic situation 
with arbitrary $m$, it is useful to focus on $m=2$. The case with $m=1$ 
can be neglected because we numerically observe that Eq.~\eqref{eq:D} vanishes in the hydrodynamic limit 
for any odd $m$. For now, let us consider $m=2$ and 
$\alpha_1=+$ and $\alpha_2=-$. One can use~\eqref{eq:idpm} to obtain 
\begin{equation}
	\label{eq:expanding}
	\mathrm{Tr}[(G^+G^-)^{q_1} G^{\alpha_1} (G^+G^-)^{q_2} G^{\alpha_{2}}]= 
	e^{-i\ell \sum_{j=1}^{2q_1+2q_2+2}k_{j-1}}\prod_{j=1}^{2q_1+2q_2+2}(e^{ik_{j-1}\ell}+s_{j}e^{ik_{j}\ell}), 
\end{equation}
where $s_j=1$ except for the sites near the positions of the defects. Precisely, 
$s_j=-1$ if a ``misplaced'' $G^-$ is inserted at $j+1$, and $s_{j}=-1$ if $G^+$ 
is inserted at $j$. Clearly, Eq.~\eqref{eq:expanding} can be generalized to account for generic  
$\alpha_l$. 
A straightforward although tedious calculation allows one to obtain that 
\begin{multline}
	\label{eq:q12}
	\mathrm{Tr}[(G^+G^-)^{q_1}G^+(G^+G^-)^{q_2}G^-]=\ell\int_{-\pi}^\pi\frac{dk}{2\pi}
	\Big\{2 (a')^{2s}+\Big[(a'-b)^{2s}+(a'+b)^{2s}-2(a')^{2s}\Big]\Theta_1(k)\\
	+\frac{1}{2}\Big[(a'-b)^{2(q_1+1)}(a'+b)^{2q_2}+(a'+b)^{2(q_1+1)}(a'-b)^{2q_2}-2(a')^{2s}\Big]\Theta_2(k)
\Big\}, 
\end{multline}
where we defined $s:=q_1+q_2+1$. 
Interestingly, the first two terms do not contain information about 
the defects. In fact they coincide with the first two terms 
in~\eqref{eq:tr-pm-d} after changing $n\to q_1+q_2+1$. They depend only on the total number of 
operators $G^\pm$ present in~\eqref{eq:D}. On the other hand, the term multiplying $\Theta_2(k)$ 
(third term in~\eqref{eq:q12}) depends on the defects. This  term is obtained from 
the second one by replacing 
$(a'-b)^{2q_1+2q_2+1}\to(a'-b)^{2q_1+2}(a'+b)^{2q_2}$ and $(a'+b)^{2q_1+2q_2+2}\to(a'+b)^{2q_1+2}(a'-b)^{2q_2}$. 
The change in the relative sign between $a'$ and $b$ reflects the presence of the 
defects in~\eqref{eq:D}. 
%
\begin{figure}[t]
\centering
\includegraphics[width=0.85\textwidth]{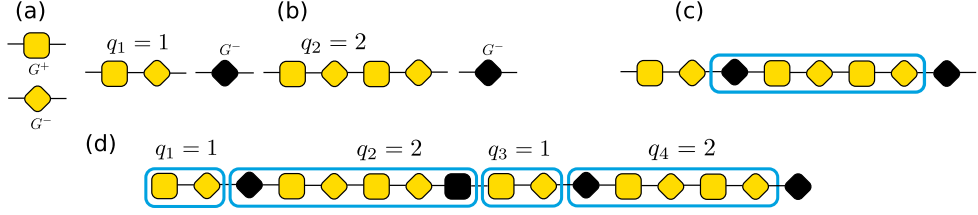}
\caption{ Pictorial illustration of the operators $G^\pm$ (a). 
 In (b) we show the pictorial definition of $(G^+G^-)^{q_1} 
 G^-(G^+G^-)^{q_2}G^-$ with $q_1=1$ and $q_2=2$. In (c) we 
 show the effect of two defects due to the insertion of two misplaced 
 operators (black symbols). The number of operators 
 between the two defects is $2q_2+d_{12}=5$. (d) The case with 
 four defects with $q_1=q_3=1$ and $q_2=q_4=2$. 
 The contribution of the configuration to the last term in~\eqref{eq:q12-gen-1} 
 is $(a'+b)^{m+2q_1+2q_3-d_{12}-d_{34}}(a'-b)^{2q_2+2q_4+d_{12}+d_{34}}$ 
 plus the term with $a'-b$ and $a'+b$ exchanged. 
}
\label{fig:string}
\end{figure}
%
A similar structue is present for generic $\alpha_1,\alpha_2$. We verified that  
\begin{multline}
	\label{eq:q12-gen}
	\mathrm{Tr}\left(\prod_{l=1}^2(G^+G^-)^{q_l}G^{\alpha_l}\right)=\ell\int_{-\pi}^\pi\frac{dk}{2\pi}
	\Big\{2 (a')^{2s}+\Big[(a'-b)^{2s}+(a'+b)^{2s}-2(a')^{2s}\Big]\Theta_1(k)\\
	+\frac{1}{2}\Big[(a'-b)^{2(q_1+1)-d_{1,2}}(a'+b)^{2q_2+d_{1,2}}+(a'+b\leftrightarrow a'-b)-2(a')^{2s}\Big]
	\Theta_2(k)
\Big\}. 
\end{multline}
Here we defined $d_{i,j}$ as
\begin{equation}
	\label{eq:d-def}
	d_{i,j}:=\left\{\begin{array}{cc}
		1 & \mathrm{for}\,\,(\alpha_i,\alpha_j)=(+,+)\\
		1 & \mathrm{for}\,\,(\alpha_i,\alpha_j)=(-,-)\\
		0 & \mathrm{for}\,\,(\alpha_i,\alpha_j)=(+,-)\\
		2 & \mathrm{for}\,\,(\alpha_i,\alpha_j)=(-,+)
	\end{array}
	\right.
\end{equation}
Let us now discuss the case with generic $m$. We verified that 
formula~\eqref{eq:q12-gen} can be generalized to arbitrary number 
of defects as 
\begin{multline}
	\label{eq:q12-gen-1}
	\mathrm{Tr}\left(\prod_{l=1}^m(G^+G^-)^{q_l}G^{\alpha_l}\right)=
	\ell\int_{-\pi}^\pi\frac{dk}{2\pi}
	\Big\{2 (a')^{2s}+\Big[(a'-b)^{2s}+(a'+b)^{2s}-2(a')^{2s}\Big]\Theta_1\\
	+\frac{1}{2}\Big[(a'-b)^{m+\sum_{l}(2q_{2l-1}-d_{2l-1,2l})}(a'+b)^{\sum_{l}(2q_{2l}+d_{2l-1,2l})}
		+(a'+b\leftrightarrow a'-b)
-2(a')^{2s}\Big]\Theta_2
\Big\}, 
\end{multline}
with $d_{i,j}$ defined in~\eqref{eq:d-def} and $s:=m/2+\sum_k q_k$. 
Again, as for the case with $m=2$ (cf.~\eqref{eq:q12}) 
the term multiplying $\Theta_1(k)$ does not depend on the defects insertions. 
Oppositely, the term multiplying $\Theta_2$ contains information about 
the defects. 
The structure of this term is illustrated in Fig.~\ref{fig:string}. In  Fig.~\ref{fig:string} (a) 
we denote with a square and a diamond the two operators $G^+$ and $G^-$, respectively. 
In (b) we show the multiplication of the 
string of operators with $m=2$, $q_1=1$, $q_2=2$ and $\alpha_1=-,\alpha_2=-$. The result is shown in 
Fig.~\ref{fig:string} (c). Defects are now present at places where the same operator is 
on consecutive sites. The box encloses the operators in between two defects. 
Notice that the number of operators in the box is $2q_2+d_{12}=5$ (cf.~Eq.~\eqref{eq:d-def}).  
Similarly, one can recover the other cases of Eq.~\eqref{eq:d-def} by considering other 
values of $\alpha_1,\alpha_2$. 
A more complicated contraction with $m=4$ is shown in Fig.~\ref{fig:string} (d). Now we have 
$\{\alpha_1,\alpha_2,\alpha_3,\alpha_4\}=\{-,+,-,-\}$ and $\{q_1,q_2,q_3,q_4\}=\{1,2,1,2\}$. 
In the last term in~\eqref{eq:q12-gen-1}, this corresponds to 
$(a'+b)^{4+2q_1+2q_3-d_{1,2}-d_{3,4}}(a-b)^{2q_2+2q_4+d_{1,2}+d_{2,4}}$ plus 
the term with the relative sign between $a'$ and $b$  exchanged. 

\begin{multline}
\label{eq:t-last}
\mathrm{Tr}\left(\prod_{l=1}^m (\mathds{1}_{2\ell}+G^+G^-)^{-1}G^{\alpha_l}\right)=\\
\ell\int_{-\pi}^\pi\frac{dk}{2\pi}
\Big\{2\Big(\frac{a'}{1+(a')^2}\Big)^m+\Big[\Big(\frac{a'-b}{1+(a'-b)^2}\Big)^m
+\Big(\frac{a'+b}{1+(a'+b)^2}\Big)^m
-2\Big(\frac{a'}{1+(a')^2}\Big)^m\Big]\Theta_1\\
+\frac{1}{2}\Big[\frac{(a'+b)^{m-\sum_{l}d_{2l-1,2l}}}{[1+(a'+b)^2]^{m/2}}
\frac{(a'-b)^{\sum_{l}d_{2l-1,2l}}}{[1+(a'-b)^2]^{m/2}}+(a'+b\leftrightarrow a'-b)
-\frac{2(a')^m}{[1+(a')^2]^m}
\Big]\Theta_2
\Big\}. 
\end{multline}
After summing over $\alpha_l=\pm$, we obtain 
\begin{multline}
\label{eq:s-last}
\mathrm{Tr}\left(\sum_{\{\alpha_l=\pm\}}\prod_{l=1}^m (\mathds{1}_{2\ell}+G^+G^-)^{-1}G^{\alpha_l}\right)=
2^{m+1}\ell\int_{-\pi}^\pi\frac{dk}{2\pi}
\Big\{\Big(\frac{a'}{1+(a')^2}\Big)^m\\+\frac{1}{2}\Big[\Big(\frac{a'-b}{1+(a'-b)^2}\Big)^m
+\Big(\frac{a'+b}{1+(a'+b)^2}\Big)^m
-2\Big(\frac{a'}{1+(a')^2}\Big)^m\Big]\Theta_1(k)\\
+\frac{1}{2}\Big[\frac{(a')^{m}}{[1+(a'+b)^2]^{m/2}[1+(a'-b)^2]^{m/2}}
-\frac{(a')^m}{[1+(a')^2]^m}
\Big]\Theta_2(k)
\Big\}. 
\end{multline}
Curiously, the term multiplying $\Theta_2(k)$ 
vanishes in the non-dissipative limit $a'\to0$, despite the fact that it shows 
the ``rise and fall'' dynamics expected for the negativity (see Fig.~\ref{fig:theta} (b)).  
From~\eqref{eq:s-last} we now obtain the moments of $G^\mathrm{T}$ (cf.~\eqref{eq:gt}) 
as 
\begin{multline}
	\label{eq:last}
	\mathrm{Tr}\Big[\big(G^{\mathrm{T}}\big)^m\Big] =
	\ell\int_{-\pi}^\pi\frac{dk}{2\pi}
	\Big\{\Big(\frac{1}{2}\pm\frac{a'}{1+(a')^2}\Big)^m\\
	+\frac{1}{2}\Big[\Big(\frac{1}{2}\pm\frac{a'-b}{1+(a'-b)^2}\Big)^m
		+\Big(\frac{1}{2}\pm\frac{a'+b}{1+(a'+b)^2}\Big)^m 
		-2\Big(\frac{1}{2}\pm\frac{a'}{1+(a')^2}\Big)^m
\Big]\Theta_1(k)\\
+\frac{1}{2}\Big[\Big(\frac{1}{2}\pm\frac{a'}{[1+(a'+b)^2]^{1/2}[1+(a'-b)^2]^{1/2}}\Big)^m
-\Big(\frac{1}{2}\pm\frac{a'}{1+(a')^2}\Big)^m
\Big]\Theta_2(k)
\Big\}, 
\end{multline}
where one has to sum over the $\pm$. Again, as for~\eqref{eq:s-last}, the term multiplying $\Theta_2(k)$ 
vanishes in the nondissipative limit.

\section{Logarithmic negativity for particle-number-conserving free-fermion systems}
\label{sec:useful}

In this appendix we report the derivation of formula~\eqref{eq:neg-comput} for the fermionic logarithmic 
negativity~\cite{shapourian2019entanglement} for free-fermion systems with fixed 
fermion number. Specifically, fermion number conservation implies 
\begin{equation}
	\label{eq:n-cons}
	\langle c_jc_l\rangle=\langle c_j^\dagger c^\dagger_l\rangle=0,\quad\forall j,l. 
\end{equation}
As a consequence of Eq.~\eqref{eq:n-cons}, the negativity can be expressed in terms of 
the correlation matrix $C_{jl}$ defined as 
\begin{equation}
	\label{eq:c-def}
	C_{jl}:=\langle c_j^\dagger c_l\rangle. 
\end{equation}
In order to show that, we start from the more general definition of the negativity in terms of 
Majorana correlation functions, which holds true also for generic, i.e., non particle-conserving 
fermion systems~\cite{shapourian2019entanglement}. Let us define the Majorana operators $a_j$ as 
\begin{equation}
	\label{eq:majo-def}
	c_j:=\frac{1}{2}(a_{2j-1}-ia_{2j}),\quad c^\dagger_j:=\frac{1}{2}(a_{2j-1}+ia_{2j}). 
\end{equation}
Here $a_j$ satisfy the standard anticommutation relations 
\begin{equation}
	\{a_j,a_l\}=2\delta_{jl}. 
\end{equation}
In the following we are going to assume that~\eqref{eq:n-cons} holds in the initial state and 
at any time. 
From the definitions~\eqref{eq:c-def} and~\eqref{eq:majo-def} we obtain that 
\begin{equation}
	\label{eq:c-1}
	C_{jl}=\frac{1}{4}(a_{2j-1}a_{2l-1}-ia_{2j-1}a_{2l}+ia_{2j}a_{2l-1}+a_{2j}a_{2l}). 
\end{equation}
After using~\eqref{eq:n-cons}, Eq.~\eqref{eq:c-1} becomes 
\begin{equation}
	C_{jl}=\frac{1}{2}(a_{2j}a_{2l}+ia_{2j}a_{2l-1}). 
\end{equation}
This implies that 
\begin{equation}
	\label{eq:gamma-c}
	2\mathrm{Re}(C_{jl})=\delta_{jl}+ia_{2j}a_{2l-1}, \quad 2i\mathrm{Im}(C_{jl})=a_{2j}a_{2l}-\delta_{jl}. 
\end{equation}
Let us now define the Majorana correlation matrix $\Gamma_{jl}$ as 
\begin{equation}
	\Gamma_{jl}:=\frac{1}{2}\langle[a_j,a_l]\rangle=\langle a_ja_l\rangle-\delta_{jl}. 
\end{equation}
From~\eqref{eq:gamma-c} we obtain that 
\begin{align}
	\label{eq:gamma-id-1}
	&\Gamma_{2j,2l}=\Gamma_{2j-1,2l-1}=2i\mathrm{Im}(C_{jl})\\
	\label{eq:gamma-id-2}
	&\Gamma_{2j-1,2l}=-\Gamma_{2j,2l-1}=-i\big(2\mathrm{Re}(C_{jl})-\delta_{jl}\big). 
\end{align}
The correlation matrix $G_{jl}$ (cf.~\eqref{eq:g-def}) is obtained 
as 
\begin{equation}
	\label{eq:G-Ga}
	G_{jl}:=2C_{jl}-\delta_{jl}=\Gamma_{2j,2l}+i\Gamma_{2j,2l-1}. 
\end{equation}
Let us now consider the $2\ell\times2\ell$ correlation matrix  $G$ restricted to subsysytem $A$, 
i.e., with $j,l\in A$ (see Fig.~\ref{fig0:bip}). We also define the $4\ell\times4\ell$ restricted 
Majorana correlation matrix as 
\begin{equation}
	\label{eq:gamma-def}
	\Gamma=\left(\begin{array}{cc}
		\Gamma_{2j,2l} & \Gamma_{2j-1,2l}\\
		-\Gamma_{2j-1,2l} & \Gamma_{2j,2l}
	\end{array}\right), 
\end{equation}
where we used~\eqref{eq:gamma-id-1} and~\eqref{eq:gamma-id-2}. 
The eigenvalues and eigenvectors of $\Gamma$ are simply related to those of 
$G_{jl}$. To show that, let us consider a generic eigenvalue $\lambda$ of 
$G_{jl}$ with eigenvector $v_j$. From~\eqref{eq:G-Ga} 
one has that 
\begin{equation}
	\label{eq:eig}
	(\Gamma_{2j,2l}+i\Gamma_{2j,2l-1})v_l=\lambda v_j,  
\end{equation}
where the sum over repeated indices is assumed. 
Now, one can check that 
\begin{equation}
	\label{eq:G-eig}
	\Gamma v_+=\lambda v_+,\quad \Gamma v_-=-\bar\lambda v_-, 
	\quad\mathrm{with}\,\, v_+=\left(\begin{array}{c}v_j\\ -iv_j\end{array}\right), 
v_-=\left(\begin{array}{c}\bar v_j\\ i\bar v_j\end{array}\right), 
\end{equation}
where $v_j$ are the components of the eigenvectors of $G_{jl}$ (cf.~\eqref{eq:eig}) and the 
bar in $\bar v_j$ denotes the complex conjugate. To verify~\eqref{eq:G-eig} one has to use that 
$\bar v_j$ satify 
\begin{equation}
	(\Gamma_{2j,2l}-i\Gamma_{2j,2l-1})\bar v_l=\bar \lambda \bar v_j, 
\end{equation}
which is obtained by taking the complex conjugate of~\eqref{eq:eig} and 
by using that $\bar \Gamma_{2j,2l}=-\Gamma_{2j,2l}$ and $\bar\Gamma_{2j,2l-1}=-\Gamma_{2j,2l-1}$. 
Furthermore, here we notice that $\lambda$ is  real, because $G_{jl}$ is an 
hermitian matrix. This means that given the eigenvalues $\lambda_k$ of $G_{jl}$, 
the eigenvalues of $\Gamma$ are organized in pairs as $(\lambda_k,-\lambda_k)$. 

A similar result holds for the matrices $\Gamma^{\pm}$ (cf.~\eqref{eq:gpm}). 
Let us first define $\Gamma^\pm$ as  
\begin{equation}
	\Gamma^\pm=\left(\begin{array}{cc}
		\Gamma^{11}& \pm i \Gamma^{12}\\
		\pm i \Gamma^{21} & -\Gamma^{22}
	\end{array}
\right). 
\end{equation}
Here $\Gamma^{pq}$ ($p,q=1,2$) are defined  in~\eqref{eq:gamma-def} with the 
constraint that $j\in A_{p}$ and $l\in A_{q}$. The eigenvalues and eigenvectors of 
$\Gamma^\pm$ are simply related to those of $G^\pm$ (cf.~\eqref{eq:gpm}) defined 
as 
\begin{equation}
	G^\pm=\left(\begin{array}{cc}
		G^{11}& \pm i G^{12}\\
		\pm i G^{21} & -G^{22}
	\end{array}
\right).
\end{equation}
First, we observe that $G^\pm$ and $\Gamma^\pm$ are not hermitian, implying 
that they have different left and right eigenvectors. 
Let us consider the right eigenvector $w^\pm_j$ ($j=1,\dots,2\ell$) 
of $G^\pm$ with eigenvalue $\mu^\pm$. 
One can show by direct computation 
that the right eigenvector $V_{\mu^\pm}$ of $\Gamma^\pm$ with eigenvalue $\mu^\pm$ is 
obtained as 
\begin{equation}
	\label{eq:eig-1}
	V_{\mu^\pm}=\big\{w^\pm_1,-iw^\pm_1,w_2^\pm,-iw_2^\pm,\dots,w_{2\ell}^\pm,-iw_{2\ell}^\pm\big\}. 
\end{equation}
The eigenvector $V_{-\bar\mu^\pm}$ associated with the other eigenvalue $-\bar\mu^\pm$ of $\Gamma^\pm$ is 
given as 
\begin{equation}
\label{eq:eig-2}
	V_{-\bar\mu^\pm}=\big\{\bar w_1^\pm,-i\bar w_1^\pm,\bar w_2^\pm,-i\bar 
		w_2^\pm,\dots,\bar w_{\ell}^\pm,-i\bar w_{\ell}^\pm,-\bar w^\pm_{\ell+1},i\bar w_{\ell+1}^\pm,
	\dots,-\bar w_{2\ell}^\pm,i\bar w_{2\ell}^\pm\big\}.
\end{equation}
Finally, in a similar way one can obtain the spectrum of $\Gamma^T$ defined 
as 
\begin{equation}
	\Gamma^T=\frac{1}{2}\left[\mathds{1}_{4\ell}-(\mathds{1}_{4\ell}+\Gamma^+\Gamma^-)^{-1}(\Gamma^++\Gamma^-)
	\right]
\end{equation}
from that of $G^T$ defined as (cf.~\eqref{eq:gt})  
\begin{equation}
	G^T=\frac{1}{2}\left[\mathds{1}_{2\ell}-(\mathds{1}_{2\ell}+G^+G^-)^{-1}(G^++G^-)\right]. 
\end{equation}
First, both $\Gamma^T$ and $G^T$ are hermitian 
matrices, and hence have real eigenvalues. Now, let us consider the 
eigenvector $Z=\{z_1,\dots,z_{2\ell}\}$ of $\mathds{1}_{2\ell}/2-G^T$ with eigenvalue $\zeta$. 
One can show that $\mathds{1}_{4\ell}/2-\Gamma^T$ has eigenvalues $\pm\zeta$. The eigenvectors 
are obtained from~\eqref{eq:eig-1} and~\eqref{eq:eig-2} after replacing $w_j\to z_j$. 
Thus, the eigenvalues $\nu_i$ of $\Gamma^T$ are given as $\nu_i=(\xi_i,1-\xi_i)$, with $\xi_i$ the 
eigenvalues of $G^T$. 
Finally, for a generic free-fermion system the negativity is obtained as~\cite{eisert2018entanglement} 
\begin{equation}
	\label{eq:app-final}
	{\cal{E}}=\frac{1}{2}\sum_{i=1}^{4\ell}\ln(\nu_i^{1/2}+(1-\nu_i)^{1/2})-\frac{1}{2}S_A^{(2)}, 
\end{equation}
where $\nu_i$ are the eigenvalues of $\Gamma^T$, and $S_A^{\scriptscriptstyle(2)}$ is the 
second R\'enyi entropy of $A=A_1\cup A_2$ (see Fig.~\ref{fig0:bip}). Given the relationship $\nu_i=(\xi_i,1-\xi_i)$ 
between $\nu_i$ and the eigenvalues $\xi_i$ of $G^T$, it is clear that~\eqref{eq:app-final} is the same as~\eqref{eq:neg-comput}.

\bibliography{bibliography.bib}

\begin{thebibliography}{10}
\providecommand{\url}[1]{\texttt{#1}}
\providecommand{\urlprefix}{URL }
\expandafter\ifx\csname urlstyle\endcsname\relax
  \providecommand{\doi}[1]{doi:\discretionary{}{}{}#1}\else
  \providecommand{\doi}{doi:\discretionary{}{}{}\begingroup
  \urlstyle{rm}\Url}\fi
\providecommand{\eprint}[2][]{\url{#2}}

\bibitem{amico2008entanglement}
L.~Amico, R.~Fazio, A.~Osterloh and V.~Vedral,
\newblock \emph{Entanglement in many-body systems},
\newblock Rev. Mod. Phys. \textbf{80}, 517 (2008),
\newblock \doi{10.1103/RevModPhys.80.517}.

\bibitem{eisert2010colloquium}
J.~Eisert, M.~Cramer and M.~B. Plenio,
\newblock \emph{Colloquium: Area laws for the entanglement entropy},
\newblock Rev. Mod. Phys. \textbf{82}, 277 (2010),
\newblock \doi{10.1103/RevModPhys.82.277}.

\bibitem{calabrese2009entanglement}
P.~Calabrese, J.~Cardy and B.~Doyon,
\newblock \emph{Entanglement entropy in extended quantum systems},
\newblock Journal of Physics A: Mathematical and Theoretical \textbf{42}(50),
  500301 (2009),
\newblock \doi{10.1088/1751-8121/42/50/500301}.

\bibitem{laflorencie2016quantum}
N.~Laflorencie,
\newblock \emph{Quantum entanglement in condensed matter systems},
\newblock Physics Reports \textbf{646}, 1 (2016),
\newblock \doi{https://doi.org/10.1016/j.physrep.2016.06.008}.

\bibitem{petruccione2002the}
H.~P. Breuer and F.~Petruccione,
\newblock \emph{The theory of open quantum systems},
\newblock Oxford University Press, Great Clarendon Street (2002).

\bibitem{rossini2021coherent}
D.~Rossini and E.~Vicari,
\newblock \emph{Coherent and dissipative dynamics at quantum phase
  transitions},
\newblock Physics Reports \textbf{936}, 1 (2021),
\newblock \doi{https://doi.org/10.1016/j.physrep.2021.08.003}.

\bibitem{alba2021spreading}
V.~Alba and F.~Carollo,
\newblock \emph{Spreading of correlations in markovian open quantum systems},
\newblock Phys. Rev. B \textbf{103}, L020302 (2021),
\newblock \doi{10.1103/PhysRevB.103.L020302}.

\bibitem{carollo2022dissipative}
F.~Carollo and V.~Alba,
\newblock \emph{Dissipative quasiparticle picture for quadratic markovian open
  quantum systems},
\newblock Phys. Rev. B \textbf{105}, 144305 (2022),
\newblock \doi{10.1103/PhysRevB.105.144305}.

\bibitem{alba2022hydrodynamics}
V.~Alba and F.~Carollo,
\newblock \emph{{Hydrodynamics of quantum entropies in {I}sing chains with
  linear dissipation}},
\newblock Journal of Physics A: Mathematical and Theoretical \textbf{55}(7),
  74002 (2022),
\newblock \doi{10.1088/1751-8121/ac48ec}.

\bibitem{calabrese2005evolution}
P.~Calabrese and J.~Cardy,
\newblock \emph{{Evolution of entanglement entropy in one-dimensional
  systems}},
\newblock Journal of Statistical Mechanics: Theory and Experiment
  \textbf{2005}(04), P04010 (2005),
\newblock \doi{10.1088/1742-5468/2005/04/p04010}.

\bibitem{fagotti2008evolution}
M.~Fagotti and P.~Calabrese,
\newblock \emph{Evolution of entanglement entropy following a quantum quench:
  Analytic results for the {XY} chain in a transverse magnetic field},
\newblock Phys. Rev. A \textbf{78}, 010306 (2008),
\newblock \doi{10.1103/PhysRevA.78.010306}.

\bibitem{alba2017entanglement}
V.~Alba and P.~Calabrese,
\newblock \emph{Entanglement and thermodynamics after a quantum quench in
  integrable systems},
\newblock Proceedings of the National Academy of Sciences \textbf{114}(30),
  7947 (2017),
\newblock \doi{10.1073/pnas.1703516114},
\newblock \eprint{https://www.pnas.org/content/114/30/7947.full.pdf}.

\bibitem{plenio2007an}
M.~B. Plenio and S.~Virmani,
\newblock \emph{An introduction to entanglement measures},
\newblock Quantum Info. Comput. \textbf{7}(1), 1 (2007).

\bibitem{lee2000partial}
J.~Lee, M.~S. Kim, Y.~J. Park and S.~Lee,
\newblock \emph{{Partial teleportation of entanglement in a noisy
  environment}},
\newblock Journal of Modern Optics \textbf{47}(12), 2151 (2000),
\newblock \doi{10.1080/09500340008235138}.

\bibitem{vidal2002computable}
G.~Vidal and R.~F. Werner,
\newblock \emph{Computable measure of entanglement},
\newblock Phys. Rev. A \textbf{65}, 032314 (2002),
\newblock \doi{10.1103/PhysRevA.65.032314}.

\bibitem{eisert2006entanglement}
J.~Eisert,
\newblock \emph{Entanglement in quantum information theory} (2006),
  \eprint{quant-ph/0610253}.

\bibitem{plenio2005logarithmic}
M.~B. Plenio,
\newblock \emph{Logarithmic negativity: A full entanglement monotone that is
  not convex},
\newblock Phys. Rev. Lett. \textbf{95}, 090503 (2005),
\newblock \doi{10.1103/PhysRevLett.95.090503}.

\bibitem{calabrese2012entanglement}
P.~Calabrese, J.~Cardy and E.~Tonni,
\newblock \emph{Entanglement negativity in quantum field theory},
\newblock Phys. Rev. Lett. \textbf{109}, 130502 (2012),
\newblock \doi{10.1103/PhysRevLett.109.130502}.

\bibitem{audenaert2002entanglement}
K.~Audenaert, J.~Eisert, M.~B. Plenio and R.~F. Werner,
\newblock \emph{Entanglement properties of the harmonic chain},
\newblock Phys. Rev. A \textbf{66}, 042327 (2002),
\newblock \doi{10.1103/PhysRevA.66.042327}.

\bibitem{eisler2015on}
V.~Eisler and Z.~Zimbor{\'{a}}s,
\newblock \emph{On the partial transpose of fermionic gaussian states},
\newblock New Journal of Physics \textbf{17}(5), 053048 (2015),
\newblock \doi{10.1088/1367-2630/17/5/053048}.

\bibitem{shapourian2019entanglement}
H.~Shapourian and S.~Ryu,
\newblock \emph{Entanglement negativity of fermions: Monotonicity, separability
  criterion, and classification of few-mode states},
\newblock Phys. Rev. A \textbf{99}, 022310 (2019),
\newblock \doi{10.1103/PhysRevA.99.022310}.

\bibitem{ruggiero2016entanglement}
P.~Ruggiero, V.~Alba and P.~Calabrese,
\newblock \emph{Entanglement negativity in random spin chains},
\newblock Phys. Rev. B \textbf{94}, 035152 (2016),
\newblock \doi{10.1103/PhysRevB.94.035152}.

\bibitem{alba2019quantum}
V.~Alba and P.~Calabrese,
\newblock \emph{Quantum information dynamics in multipartite integrable
  systems},
\newblock {EPL} (Europhysics Letters) \textbf{126}(6), 60001 (2019),
\newblock \doi{10.1209/0295-5075/126/60001}.

\bibitem{alba2017quench}
V.~Alba and P.~Calabrese,
\newblock \emph{Quench action and {R\'e}nyi entropies in integrable systems},
\newblock Phys. Rev. B \textbf{96}, 115421 (2017),
\newblock \doi{10.1103/PhysRevB.96.115421}.

\bibitem{alba2017renyi}
V.~Alba and P.~Calabrese,
\newblock \emph{R{\'{e}}nyi entropies after releasing the {N{\'{e}}el} state in
  {the XXZ spin}-chain},
\newblock Journal of Statistical Mechanics: Theory and Experiment
  \textbf{2017}(11), 113105 (2017),
\newblock \doi{10.1088/1742-5468/aa934c}.

\bibitem{mestyan2018renyi}
M.~Mesty{\'{a}}n, V.~Alba and P.~Calabrese,
\newblock \emph{R{\'{e}}nyi entropies of generic thermodynamic macrostates in
  integrable systems},
\newblock Journal of Statistical Mechanics: Theory and Experiment
  \textbf{2018}(8), 083104 (2018),
\newblock \doi{10.1088/1742-5468/aad6b9}.

\bibitem{alba2019towards}
V.~Alba,
\newblock \emph{Towards a generalized hydrodynamics description of {R\'e}nyi
  entropies in integrable systems},
\newblock Phys. Rev. B \textbf{99}, 045150 (2019),
\newblock \doi{10.1103/PhysRevB.99.045150}.

\bibitem{klobas2021entanglement}
K.~Klobas and B.~Bertini,
\newblock \emph{{Entanglement dynamics in Rule 54: Exact results and
  quasiparticle picture}},
\newblock SciPost Phys. \textbf{11}, 107 (2021),
\newblock \doi{10.21468/SciPostPhys.11.6.107}.

\bibitem{bertini2022growth}
B.~Bertini, K.~Klobas, V.~Alba, G.~Lagnese and P.~Calabrese,
\newblock \emph{{Growth of R\'enyi Entropies in Interacting Integrable Models
  and the Breakdown of the Quasiparticle Picture}},
\newblock \doi{10.48550/ARXIV.2203.17264} (2022).

\bibitem{bertini2022entanglement}
B.~Bertini, K.~Klobas and T.-C. Lu,
\newblock \emph{Entanglement negativity and mutual information after a quantum
  quench: Exact link from space-time duality},
\newblock Phys. Rev. Lett. \textbf{129}, 140503 (2022),
\newblock \doi{10.1103/PhysRevLett.129.140503}.

\bibitem{kudler2021the}
J.~Kudler-Flam, Y.~Kusuki and S.~Ryu,
\newblock \emph{The quasi-particle picture and its breakdown after local
  quenches: mutual information, negativity, and reflected entropy},
\newblock Journal of High Energy Physics \textbf{2021}(3), 146 (2021),
\newblock \doi{10.1007/JHEP03(2021)146}.

\bibitem{calabrese2012quantumquench}
P.~Calabrese, F.~H.~L. Essler and M.~Fagotti,
\newblock \emph{{Quantum quenches in the transverse field Ising chain: II.
  Stationary state properties}},
\newblock Journal of Statistical Mechanics: Theory and Experiment
  \textbf{2012}(07), P07022 (2012),
\newblock \doi{10.1088/1742-5468/2012/07/p07022}.

\bibitem{peschel2009reduced}
I.~Peschel and V.~Eisler,
\newblock \emph{Reduced density matrices and entanglement entropy in free
  lattice models},
\newblock Journal of physics A: Mathematical and Theoretical \textbf{42}(50),
  504003 (2009).

\bibitem{landi2022nonequilibrium}
G.~T. Landi, D.~Poletti and G.~Schaller,
\newblock \emph{Nonequilibrium boundary-driven quantum systems: Models,
  methods, and properties},
\newblock Rev. Mod. Phys. \textbf{94}, 045006 (2022),
\newblock \doi{10.1103/RevModPhys.94.045006}.

\bibitem{alba2021generalized}
V.~Alba, B.~Bertini, M.~Fagotti, L.~Piroli and P.~Ruggiero,
\newblock \emph{Generalized-hydrodynamic approach to inhomogeneous quenches:
  correlations, entanglement and quantum effects},
\newblock Journal of Statistical Mechanics: Theory and Experiment
  \textbf{2021}(11), 114004 (2021),
\newblock \doi{10.1088/1742-5468/ac257d}.

\bibitem{prosen2008third}
T.~Prosen,
\newblock \emph{{Third quantization: a general method to solve master equations
  for quadratic open Fermi systems}},
\newblock New Journal of Physics \textbf{10}(4), 43026 (2008),
\newblock \doi{10.1088/1367-2630/10/4/043026}.

\bibitem{alba2018entanglement}
V.~Alba and P.~Calabrese,
\newblock \emph{{Entanglement dynamics after quantum quenches in generic
  integrable systems}},
\newblock SciPost Phys. \textbf{4}, 17 (2018),
\newblock \doi{10.21468/SciPostPhys.4.3.017}.

\bibitem{starchl2022relaxation}
E.~Starchl and L.~M. Sieberer,
\newblock \emph{{Relaxation to a parity-time symmetric generalized Gibbs
  ensemble after a quantum quench in a driven-dissipative Kitaev chain}},
\newblock \doi{10.48550/ARXIV.2203.14589} (2022).

\bibitem{polkovnikov2011colloquium}
A.~Polkovnikov, K.~Sengupta, A.~Silva and M.~Vengalattore,
\newblock \emph{Colloquium: Nonequilibrium dynamics of closed interacting
  quantum systems},
\newblock Rev. Mod. Phys. \textbf{83}, 863 (2011),
\newblock \doi{10.1103/RevModPhys.83.863}.

\bibitem{calabrese2016introduction}
P.~Calabrese, F.~H.~L. Essler and G.~Mussardo,
\newblock \emph{{Introduction to `Quantum Integrability in Out of Equilibrium
  Systems'}},
\newblock Journal of Statistical Mechanics: Theory and Experiment
  \textbf{2016}(6), 064001 (2016),
\newblock \doi{10.1088/1742-5468/2016/06/064001}.

\bibitem{essler2016quench}
F.~H.~L. Essler and M.~Fagotti,
\newblock \emph{Quench dynamics and relaxation in isolated integrable quantum
  spin chains},
\newblock Journal of Statistical Mechanics: Theory and Experiment
  \textbf{2016}(6), 064002 (2016),
\newblock \doi{10.1088/1742-5468/2016/06/064002}.

\bibitem{vidmar2016generalized}
L.~Vidmar and M.~Rigol,
\newblock \emph{{Generalized Gibbs ensemble in integrable lattice models}},
\newblock Journal of Statistical Mechanics: Theory and Experiment
  \textbf{2016}(6), 064007 (2016),
\newblock \doi{10.1088/1742-5468/2016/06/064007}.

\bibitem{caux2013time}
J.-S. Caux and F.~H.~L. Essler,
\newblock \emph{Time evolution of local observables after quenching to an
  integrable model},
\newblock Phys. Rev. Lett. \textbf{110}, 257203 (2013),
\newblock \doi{10.1103/PhysRevLett.110.257203}.

\bibitem{caux2016the}
J.-S. Caux,
\newblock \emph{{The Quench Action}},
\newblock Journal of Statistical Mechanics: Theory and Experiment
  \textbf{2016}(6), 064006 (2016),
\newblock \doi{10.1088/1742-5468/2016/06/064006}.

\bibitem{coser2015partial}
A.~Coser, E.~Tonni and P.~Calabrese,
\newblock \emph{{Partial transpose of two disjoint blocks in XY spin chains}},
\newblock Journal of Statistical Mechanics: Theory and Experiment
  \textbf{2015}(8), P08005 (2015),
\newblock \doi{10.1088/1742-5468/2015/08/P08005}.

\bibitem{coser2016towards}
A.~Coser, E.~Tonni and P.~Calabrese,
\newblock \emph{Towards the entanglement negativity of two disjoint intervals
  for a one dimensional free fermion},
\newblock Journal of Statistical Mechanics: Theory and Experiment
  \textbf{2016}(3), 033116 (2016),
\newblock \doi{10.1088/1742-5468/2016/03/033116}.

\bibitem{herzog2016estimation}
C.~P. Herzog and Y.~Wang,
\newblock \emph{Estimation for entanglement negativity of free fermions},
\newblock Journal of Statistical Mechanics: Theory and Experiment
  \textbf{2016}(7), 073102 (2016),
\newblock \doi{10.1088/1742-5468/2016/07/073102}.

\bibitem{shapourian2017many}
H.~Shapourian, K.~Shiozaki and S.~Ryu,
\newblock \emph{Many-body topological invariants for fermionic
  symmetry-protected topological phases},
\newblock Phys. Rev. Lett. \textbf{118}, 216402 (2017),
\newblock \doi{10.1103/PhysRevLett.118.216402}.

\bibitem{shapourian2017partial}
H.~Shapourian, K.~Shiozaki and S.~Ryu,
\newblock \emph{Partial time-reversal transformation and entanglement
  negativity in fermionic systems},
\newblock Phys. Rev. B \textbf{95}, 165101 (2017),
\newblock \doi{10.1103/PhysRevB.95.165101}.

\bibitem{eisert2018entanglement}
J.~Eisert, V.~Eisler and Z.~Zimbor\'as,
\newblock \emph{Entanglement negativity bounds for fermionic gaussian states},
\newblock Phys. Rev. B \textbf{97}, 165123 (2018),
\newblock \doi{10.1103/PhysRevB.97.165123}.

\bibitem{shapourian2019finite}
H.~Shapourian and S.~Ryu,
\newblock \emph{{Finite-temperature entanglement negativity of free fermions}},
\newblock Journal of Statistical Mechanics: Theory and Experiment
  \textbf{2019}(4), 43106 (2019),
\newblock \doi{10.1088/1742-5468/ab11e0}.

\bibitem{eisert2016entanglement}
J.~Eisert, V.~Eisler and Z.~Zimbor\'as,
\newblock \emph{Entanglement negativity bounds for fermionic gaussian states},
\newblock Phys. Rev. B \textbf{97}, 165123 (2018),
\newblock \doi{10.1103/PhysRevB.97.165123}.

\bibitem{ferraro2008thermal}
A.~Ferraro, D.~Cavalcanti, A.~Garc\'{\i}a-Saez and A.~Ac\'{\i}n,
\newblock \emph{Thermal bound entanglement in macroscopic systems and area
  law},
\newblock Phys. Rev. Lett. \textbf{100}, 080502 (2008),
\newblock \doi{10.1103/PhysRevLett.100.080502}.

\bibitem{anders2008entanglement}
J.~Anders and A.~Winter,
\newblock \emph{{Entanglement and Separability of Quantum Harmonic Oscillator
  Systems at Finite Temperature}},
\newblock Quantum Info. Comput. \textbf{8}(3), 245 (2008).

\bibitem{marcovitch2009critical}
S.~Marcovitch, A.~Retzker, M.~B. Plenio and B.~Reznik,
\newblock \emph{{Critical and noncritical long-range entanglement in
  Klein-Gordon fields}},
\newblock Phys. Rev. A \textbf{80}, 012325 (2009),
\newblock \doi{10.1103/PhysRevA.80.012325}.

\bibitem{eisler2014entanglement}
V.~Eisler and Z.~Zimbor{\'{a}}s,
\newblock \emph{{Entanglement negativity in the harmonic chain out of
  equilibrium}},
\newblock New Journal of Physics \textbf{16}(12), 123020 (2014),
\newblock \doi{10.1088/1367-2630/16/12/123020}.

\bibitem{sherman2016nonzero}
N.~E. Sherman, T.~Devakul, M.~B. Hastings and R.~R.~P. Singh,
\newblock \emph{Nonzero-temperature entanglement negativity of quantum spin
  models: Area law, linked cluster expansions, and sudden death},
\newblock Phys. Rev. E \textbf{93}, 022128 (2016),
\newblock \doi{10.1103/PhysRevE.93.022128}.

\bibitem{denobili2016entanglement}
C.~D. Nobili, A.~Coser and E.~Tonni,
\newblock \emph{{Entanglement negativity in a two dimensional harmonic lattice:
  area law and corner contributions}},
\newblock Journal of Statistical Mechanics: Theory and Experiment
  \textbf{2016}(8), 83102 (2016),
\newblock \doi{10.1088/1742-5468/2016/08/083102}.

\bibitem{alba2013entanglement}
V.~Alba,
\newblock \emph{{Entanglement negativity and conformal field theory: A Monte
  Carlo study}},
\newblock Journal of Statistical Mechanics: Theory and Experiment
  \textbf{2013}(5) (2013),
\newblock \doi{10.1088/1742-5468/2013/05/P05013}.

\bibitem{chung2014entanglement}
C.-M. Chung, V.~Alba, L.~Bonnes, P.~Chen and A.~M. L\"auchli,
\newblock \emph{{Entanglement negativity via the replica trick: A quantum Monte
  Carlo approach}},
\newblock Phys. Rev. B \textbf{90}, 064401 (2014),
\newblock \doi{10.1103/PhysRevB.90.064401}.

\bibitem{calabrese2013entanglement}
P.~Calabrese, L.~Tagliacozzo and E.~Tonni,
\newblock \emph{{Entanglement negativity in the critical Ising chain}},
\newblock Journal of Statistical Mechanics: Theory and Experiment
  \textbf{2013}(05), P05002 (2013),
\newblock \doi{10.1088/1742-5468/2013/05/p05002}.

\bibitem{wu2020entanglement}
K.-H. Wu, T.-C. Lu, C.-M. Chung, Y.-J. Kao and T.~Grover,
\newblock \emph{{Entanglement R\'enyi Negativity across a Finite Temperature
  Transition: A Monte Carlo Study}},
\newblock Phys. Rev. Lett. \textbf{125}, 140603 (2020),
\newblock \doi{10.1103/PhysRevLett.125.140603}.

\bibitem{lu2020entanglement}
T.-C. Lu and T.~Grover,
\newblock \emph{Entanglement transitions as a probe of quasiparticles and
  quantum thermalization},
\newblock Phys. Rev. B \textbf{102}, 235110 (2020),
\newblock \doi{10.1103/PhysRevB.102.235110}.

\bibitem{wichterich2009scaling}
H.~Wichterich, J.~Molina-Vilaplana and S.~Bose,
\newblock \emph{Scaling of entanglement between separated blocks in spin chains
  at criticality},
\newblock Phys. Rev. A \textbf{80}, 010304 (2009),
\newblock \doi{10.1103/PhysRevA.80.010304}.

\bibitem{wichterich2010universality}
H.~Wichterich, J.~Vidal and S.~Bose,
\newblock \emph{{Universality of the negativity in the Lipkin-Meshkov-Glick
  model}},
\newblock Phys. Rev. A \textbf{81}, 032311 (2010),
\newblock \doi{10.1103/PhysRevA.81.032311}.

\bibitem{bayat2012entanglement}
A.~Bayat, S.~Bose, P.~Sodano and H.~Johannesson,
\newblock \emph{{Entanglement Probe of Two-Impurity Kondo Physics in a Spin
  Chain}},
\newblock Phys. Rev. Lett. \textbf{109}, 066403 (2012),
\newblock \doi{10.1103/PhysRevLett.109.066403}.

\bibitem{turkeshi2020entanglement}
X.~Turkeshi, P.~Ruggiero, V.~Alba and P.~Calabrese,
\newblock \emph{Entanglement equipartition in critical random spin chains},
\newblock Phys. Rev. B \textbf{102}, 014455 (2020),
\newblock \doi{10.1103/PhysRevB.102.014455}.

\bibitem{mbeng2017negativity}
G.~Mbeng, V.~Alba and P.~Calabrese,
\newblock \emph{{Negativity spectrum in 1D gapped phases of matter}},
\newblock Journal of Physics A: Mathematical and Theoretical \textbf{50}(19)
  (2017),
\newblock \doi{10.1088/1751-8121/aa6734}.

\bibitem{wald2020entanglement}
S.~Wald, R.~Arias and V.~Alba,
\newblock \emph{{Entanglement and classical fluctuations at finite-temperature
  critical points}},
\newblock Journal of Statistical Mechanics: Theory and Experiment
  \textbf{2020}(3) (2020),
\newblock \doi{10.1088/1742-5468/ab6b19}.

\bibitem{shapourian2021entanglement}
H.~Shapourian, S.~Liu, J.~Kudler-Flam and A.~Vishwanath,
\newblock \emph{Entanglement negativity spectrum of random mixed states: A
  diagrammatic approach},
\newblock PRX Quantum \textbf{2}, 030347 (2021),
\newblock \doi{10.1103/PRXQuantum.2.030347}.

\bibitem{calabrese2013entanglementnegativity}
P.~Calabrese, J.~Cardy and E.~Tonni,
\newblock \emph{{Entanglement negativity in extended systems: a field
  theoretical approach}},
\newblock Journal of Statistical Mechanics: Theory and Experiment
  \textbf{2013}(02), P02008 (2013),
\newblock \doi{10.1088/1742-5468/2013/02/p02008}.

\bibitem{calabrese2014finite}
P.~Calabrese, J.~Cardy and E.~Tonni,
\newblock \emph{{Finite temperature entanglement negativity in conformal field
  theory}},
\newblock Journal of Physics A: Mathematical and Theoretical \textbf{48}(1),
  15006 (2014),
\newblock \doi{10.1088/1751-8113/48/1/015006}.

\bibitem{wen2015entanglement}
X.~Wen, P.-Y. Chang and S.~Ryu,
\newblock \emph{Entanglement negativity after a local quantum quench in
  conformal field theories},
\newblock Phys. Rev. B \textbf{92}, 075109 (2015),
\newblock \doi{10.1103/PhysRevB.92.075109}.

\bibitem{ruggiero2016negativity}
P.~Ruggiero, V.~Alba and P.~Calabrese,
\newblock \emph{Negativity spectrum of one-dimensional conformal field
  theories},
\newblock Phys. Rev. B \textbf{94}, 195121 (2016),
\newblock \doi{10.1103/PhysRevB.94.195121}.

\bibitem{alba2017entanglementspectrum}
V.~Alba, P.~Calabrese and E.~Tonni,
\newblock \emph{{Entanglement spectrum degeneracy and the Cardy formula in
  $1+1$ dimensional conformal field theories}},
\newblock Journal of Physics A: Mathematical and Theoretical \textbf{51}(2),
  24001 (2017),
\newblock \doi{10.1088/1751-8121/aa9365}.

\bibitem{cornfeld2019measuring}
E.~Cornfeld, E.~Sela and M.~Goldstein,
\newblock \emph{Measuring fermionic entanglement: Entropy, negativity, and spin
  structure},
\newblock Phys. Rev. A \textbf{99}, 062309 (2019),
\newblock \doi{10.1103/PhysRevA.99.062309}.

\bibitem{kudler-flam2020the}
J.~Kudler-Flam, H.~Shapourian and S.~Ryu,
\newblock \emph{{The negativity contour: a quasi-local measure of entanglement
  for mixed states}},
\newblock SciPost Phys. \textbf{8}, 63 (2020),
\newblock \doi{10.21468/SciPostPhys.8.4.063}.

\bibitem{coser2014entanglement}
A.~Coser, E.~Tonni and P.~Calabrese,
\newblock \emph{{Entanglement negativity after a global quantum quench}},
\newblock Journal of Statistical Mechanics: Theory and Experiment
  \textbf{2014}(12), P12017 (2014),
\newblock \doi{10.1088/1742-5468/2014/12/p12017}.

\bibitem{feldman2019dynamics}
N.~Feldman and M.~Goldstein,
\newblock \emph{Dynamics of charge-resolved entanglement after a local quench},
\newblock Phys. Rev. B \textbf{100}, 235146 (2019),
\newblock \doi{10.1103/PhysRevB.100.235146}.

\bibitem{kudler2020correlation}
J.~Kudler-Flam, Y.~Kusuki and S.~Ryu,
\newblock \emph{Correlation measures and the entanglement wedge cross-section
  after quantum quenches in two-dimensional conformal field theories},
\newblock Journal of High Energy Physics \textbf{2020}(4), 74 (2020),
\newblock \doi{10.1007/JHEP04(2020)074}.

\bibitem{gruber2020time}
M.~Gruber and V.~Eisler,
\newblock \emph{{Time evolution of entanglement negativity across a defect}},
\newblock Journal of Physics A: Mathematical and Theoretical \textbf{53}(20),
  205301 (2020),
\newblock \doi{10.1088/1751-8121/ab831c}.

\bibitem{elben2020mixed}
A.~Elben, R.~Kueng, H.-Y.~R. Huang, R.~van Bijnen, C.~Kokail, M.~Dalmonte,
  P.~Calabrese, B.~Kraus, J.~Preskill, P.~Zoller and B.~Vermersch,
\newblock \emph{Mixed-state entanglement from local randomized measurements},
\newblock Phys. Rev. Lett. \textbf{125}, 200501 (2020),
\newblock \doi{10.1103/PhysRevLett.125.200501}.

\bibitem{murciano2021quench}
S.~Murciano, V.~Alba and P.~Calabrese,
\newblock \emph{Quench dynamics of {R}\'enyi negativities and the quasiparticle
  picture} (2021), \eprint{arXiv:2110.14589}.

\bibitem{parez2022dynamics}
G.~Parez, R.~Bonsignori and P.~Calabrese,
\newblock \emph{Dynamics of charge-imbalance-resolved entanglement negativity
  after a quench in a free-fermion model} (2022), \eprint{arXiv:2202.05309}.

\bibitem{wolf2008area}
M.~M. Wolf, F.~Verstraete, M.~B. Hastings and J.~I. Cirac,
\newblock \emph{Area laws in quantum systems: Mutual information and
  correlations},
\newblock Phys. Rev. Lett. \textbf{100}, 070502 (2008),
\newblock \doi{10.1103/PhysRevLett.100.070502}.

\bibitem{calabrese2012quantum}
P.~Calabrese, F.~H.~L. Essler and M.~Fagotti,
\newblock \emph{{Quantum quench in the transverse field Ising chain: I. Time
  evolution of order parameter correlators}},
\newblock Journal of Statistical Mechanics: Theory and Experiment
  \textbf{2012}(07), P07016 (2012),
\newblock \doi{10.1088/1742-5468/2012/07/p07016}.

\bibitem{wong2001asymptotic}
R.~Wong,
\newblock \emph{{Asymptotic Approximations of Integrals}},
\newblock Society for Industrial and Applied Mathematics,
\newblock \doi{10.1137/1.9780898719260} (2001).

\bibitem{caceffo2022entanglement}
F.~Caceffo and V.~Alba,
\newblock \emph{Entanglement negativity in a fermionic chain with dissipative
  defects: Exact results}  (2022),
\newblock \doi{10.48550/ARXIV.2209.14164}.

\bibitem{parez2021exact}
G.~Parez, R.~Bonsignori and P.~Calabrese,
\newblock \emph{{Exact quench dynamics of symmetry resolved entanglement in a
  free fermion chain}},
\newblock Journal of Statistical Mechanics: Theory and Experiment
  \textbf{2021}(9), 93102 (2021),
\newblock \doi{10.1088/1742-5468/ac21d7}.

\bibitem{alba2021unbounded}
V.~Alba,
\newblock \emph{{Unbounded entanglement production via a dissipative
  impurity}},
\newblock SciPost Phys. \textbf{12}, 11 (2022),
\newblock \doi{10.21468/SciPostPhys.12.1.011}.

\end{thebibliography}
\nolinenumbers

\end{document}